\documentclass[preprint,3p,12pt]{elsarticle}

\journal{Annals of Physics (N.Y.)}
\usepackage{amssymb}
\usepackage{amsmath}
\usepackage{array}

\newcommand{\ba}{\begin{eqnarray}}
\newcommand{\ea}{\end{eqnarray}}
\newcommand{\bmath}{\begin{mathletters}}
\newcommand{\emath}{\end{mathletters}}
\newcommand{\ban}{\begin{eqnarray*}}
\newcommand{\ean}{\end{eqnarray*}}

\newcommand{\bsub}{\begin{subequations}}
\newcommand{\esub}{\end{subequations}}

\begin{document}

\begin{frontmatter}

\title{Quantum phase transitions in Bose-Fermi systems}
\author[yale]{D. Petrellis}
\ead{dimitris.petrellis@yale.edu}

\author[huji]{A. Leviatan\corref{cor1}}
\ead{ami@phys.huji.ac.il}

\author[yale]{F. Iachello}
\ead{francesco.iachello@yale.edu}

\cortext[cor1]{Corresponding author.}

\address[yale]{Center for Theoretical Physics, Sloane Physics Laboratory, 
Yale University, New Haven, Connecticut~06520-8120, USA}
\address[huji]{Racah Institute of Physics, The Hebrew University, 
Jerusalem 91904, Israel}

\begin{abstract}
Quantum phase transitions in a system of $N$ bosons with angular momentum 
$L=0,2$ (s,d) and a single fermion with angular momentum~$j$ are investigated
both classically and quantum mechanically. It is shown that the presence of
the odd fermion strongly influences the location and nature of the phase
transition, especially the critical value of the control parameter at which
the phase transition occurs. Experimental evidence for the U(5)-SU(3) 
(spherical to axially-deformed) transition in odd-even nuclei is presented.
\end{abstract}

\begin{keyword}
Bose-Fermi systems; Algebraic models; Quantum shape-phase transitions;\\
Interacting boson-fermion model (IBFM)

\vspace{12pt}
\PACS 21.60.Fw; 21.60.Ev; 05.30.Rt
\end{keyword}

\end{frontmatter}

\section{Introduction}
\label{Sec1}

Quantum phase transitions (QPT) are qualitative changes in the structure of
a physical system induced by a change in one or more parameters that appear
in the quantum Hamiltonian describing the system. Originally introduced in
nuclear physics~\cite{gilmore1,gilmore2}, where they were called
ground state phase transitions, they have received recently considerable
attention in condensed matter physics and other areas~\cite{vojta}. Quantum
phase transitions were investigated both classically~\cite{diep,feng}
and quantum mechanically~\cite{iac-scholten} in even-even nuclei in the
early 1980's within the framework of the Interacting Boson\ Model (IBM), a
model of nuclei in terms of correlated pairs of nucleons with $L=0,2$
treated as bosons ($s,d$ bosons)~\cite{iac111}. In recent years, this study
has been greatly expanded, including the aspect of symmetries (critical point 
symmetries~\cite{cps1,cps2,cps3}, quasidynamical~\cite{qds} and 
partial dynamical symmetries~\cite{pds}) and their empirical 
evidence~\cite{exp1,exp2}, 
the study of finite-N effects~\cite{levgin03,zamfir,lev05,lev06} and of the 
corresponding scaling behavior~\cite{rowe04,dusuel05,williams,werner}. 
The latter plays 
an important role in nuclei, since the nucleus is composed of a finite
number of particles, $N$, and phase transitions, i.e. discontinuities in
some quantities, are defined only in the limit $N\rightarrow \infty $. Also, 
the concept of QPTs has been enlarged to include excited states quantum 
phase transitions (ESQPT)~\cite{caprio}, that is qualitative changes in the
structure of a physical system as a function of excitation energy. Quantum
phase transitions in the IBM are particularly interesting since this model
has an algebraic structure, U(6), which is rather complex, giving rise
within its parameter space to both first and second order transitions. On
the other side, because of its algebraic structure, one can work out the
``phases" in explicit analytic form, since these correspond to dynamic
symmetries of the Hamiltonian. This situation is summarized in several 
review papers~\cite{cejnar09,cejnar10,iaccap10}, where a complete list of 
references is given. 

In this article we address the problem of how QPTs are affected by the
presence of fermions in addition to bosons (QPTs in Bose-Fermi systems,
odd-even nuclei). Nuclei offer a unique opportunity to study this problem
since one has a model, the 
Interacting Boson-Fermion Model (IBFM)~\cite{iac1}, 
in terms of correlated pairs with $L=0,2$ ($s,d$ bosons) 
and unpaired particles with angular momentum~$j$ ($j$~fermions),
where this problem can be addressed in explicit form. (The problem is also
of interest to the condensed matter physics community where one is
interested in the motion of fermionic impurities in a bath of 
bosons~\cite{lehur}.) 
Studies of QPTs in odd-even nuclei were implicitly initiated years
ago by Scholten \cite{scholten-blasi}. Several explicit studies have 
recently been made by Alonso \textit{et al.}~\cite{alonso1,alonso2,alonso3} 
and by B\"oy\"ukata \textit{et al.}~\cite{boy10}, 
who also have suggested a simple form of the IBFM
Hamiltonian particularly well suited to study QPTs in odd-even nuclei
because of its supersymmetric properties. Here we further expand on these
studies and present the general theory of a single fermion with angular
momentum~$j$ interacting with a system of $s,d$ bosons (pairs with angular
momentum $L=0,2$) and study QPTs in this system both classically and quantum
mechanically. We consider specifically the case of a particle with $j=11/2$.
After introducing the model Hamiltonian in Section~\ref{Sec2}, 
we present in Section~\ref{Sec3} a classical analysis with novel results 
on single particle levels in a deformed field with both $\beta $ and 
$\gamma $ deformation of interest not only to the algebraic description 
but also to its geometric counterpart. 
In Section~\ref{Sec4} we present a quantal analysis of the same problem, 
introduce correlation diagrams for Bose-Fermi systems and study the 
classical-quantal correspondence.

QPTs in odd-even nuclei would be a mere academic (and very complex) exercise
were not for the fact that there is experimental evidence for the occurrence
of these phase transitions in the odd-proton nuclei $_{61}$Pm, $_{63}$Eu and 
$_{65}$Tb. In the last section (Section~\ref{Sec5}), 
we discuss this evidence and
compare the data with realistic IBFM calculations. The evidence is
particularly clear in the unique parity negative parity levels originating
from the $h_{11/2}$ level ($\ell =5,j=11/2$) and we therefore study these
states. The large value of~$j$ has also the advantage that it provides a way
to unravel the many complications of QPTs in Bose-Fermi systems which are not
present for small values of~$j$, $j=1/2$ being trivial and $j=3/2$ being
analytically solvable.

\section{Model Hamiltonian}
\label{Sec2}

We consider here the model Hamiltonian of a system of $N$ bosons with
angular momentum $L=0,2$ ($s,d$ bosons) coupled to a fermion with angular
momentum,~$j$, as exemplified in the IBFM~\cite{iac1}, 

\begin{equation}
H=H_{B}+H_{F}+V_{BF}
\label{Eq:hBF}
\end{equation}
with
\begin{eqnarray}
H_{B} &=&\varepsilon_{0}\left[ \left( 1-\xi \right) \hat{n}_{d}-\frac{\xi }{
4N}\hat{Q}^{\chi }\cdot \hat{Q}^{\chi }\right]  
\nonumber \\
H_{F} &=&\varepsilon _{j}\,\hat{n}_{j}  
\nonumber \\
V_{BF} &=&V_{BF}^{MON}+V_{BF}^{QUAD}+V_{BF}^{EXC} ~.
\label{Eq:hBhFvBF}
\end{eqnarray}
Here the subscripts $B$, $F$ and $BF$ refer to the boson, fermion, 
and boson-fermion parts of the full Hamiltonian and 
\begin{eqnarray}
V_{BF}^{MON} &=&A\,\hat{n}_{d}\,\hat{n}_{j}  
\nonumber \\
V_{BF}^{QUAD} &=&\Gamma\, \hat{Q}^{\chi }\cdot \hat{q}_{j}  
\nonumber \\
V_{BF}^{EXC} &=&\Lambda\, \sqrt{2j+1}:[ ( d^{\dag }\times \tilde{a}_{j})^{(j)}
\times ( \tilde{d}\times a_{j}^{\dag })^{(j)}]^{(0)}: ~.
\label{Eq:MON}
\end{eqnarray}
The superscripts $MON$, $QUAD$ and $EXC$ label the monopole, quadrupole 
and exchange terms, respectively. Their coefficients are 
$A\equiv -A_{j}/\sqrt{5(2j+1)}$, $\Gamma \equiv \Gamma _{jj}/\sqrt{5}$, 
and $\Lambda \equiv \Lambda _{jj}^{j}/\sqrt{2j+1}$ 
in the notation of~\cite{iac1}, and 
\begin{eqnarray}
\hat{n}_{d} &=& d^{\dag }\cdot \tilde{d}  
\nonumber \\
\hat{Q}^{\chi } &=& ( d^{\dag }\times s+s^{\dag }\times \tilde{d})^{(2)} 
+\chi ( d^{\dag }\times \tilde{d}) ^{(2)}  
\nonumber \\
\hat{n}_{j} &=&  -\sqrt{2j+1} ( a_{j}^{\dag }\times\tilde{a}_{j})^{(0)}  
\nonumber \\
\hat{q}_{j} &=& ( a_{j}^{\dag }\times \tilde{a}_{j} )^{(2)} ~.
\label{Eq:operators}
\end{eqnarray}
In these formulas, dots $\cdot $\ denote scalar products, symbols $\times$
denote tensor products and $:$ denotes normal ordering. Also, 
$s^{\dag},d_{\mu }^{\dag }$ ($s,d_{\mu }$) ($\mu =0,\pm 1,\pm 2$) denote 
creation (annihilation) operators for $s,d$ bosons and 
$a_{j,m}^{\dag }$($a_{j,m}$)\ 
($m=\pm \frac{1}{2},\pm\frac{3}{2},...,\pm j$) 
creation (annihilation) operators for fermions
with angular momentum~$j$. The adjoint operators are $\tilde{d}_{\mu
}=(-)^{\mu }d_{-\mu },\tilde{a}_{j,m}=(-)^{j-m}a_{j,-m}$.

\begin{figure}[t!]
\centering
\includegraphics[scale=0.7]{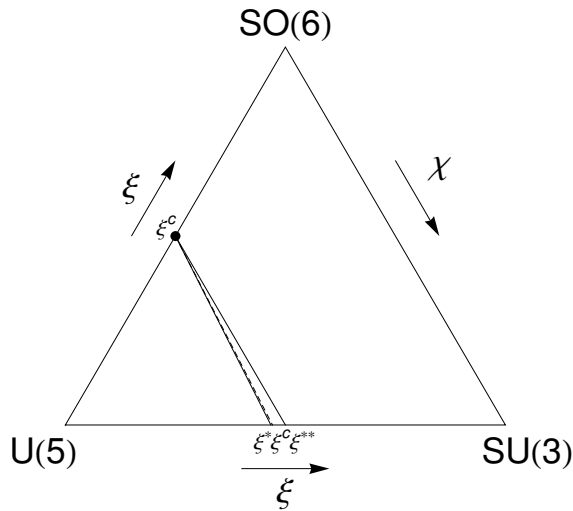}
\caption{Phase diagram of a system of $s,d$ bosons. The three phases are
denoted by their symmetry, U(5), SU(3) and SO(6). A line of first order
transitions ends in a point of second order transition 
in-between the U(5) and SO(6) limits. The spinodal,
critical and antispinodal points are denoted by $\xi ^{\ast },\xi ^{c},\xi
^{\ast \ast }$ respectively.}
\label{fig1}
\end{figure}

The quantum phase transitions of the boson part of the 
Hamiltonian~(\ref{Eq:hBhFvBF}) are very well known~\cite{reviews1}. 
There are three ``phases", characterized by their symmetry, 
U(5), SU(3) and SO(6), and two control parameters, $\xi $ and $\chi $. 
We consider here the range of parameters, 
$0\leq \xi \leq 1$, $-\frac{\sqrt{7}}{2}\leq \chi\leq 0$. 
As $\xi$ changes from $0$ to $1$ the system undergoes a quantum phase
transition. No phase transition occurs when $\chi $ changes from $0$ to 
$-\frac{\sqrt{7}}{2}$. The phase transition as a function of $\xi $ is 
first order from U(5) to SU(3) (spherical to axially deformed phases) 
and second order from U(5) to SO(6) 
(spherical to $\gamma $-unstable deformed phases). The situation is
summarized in Fig.~\ref{fig1}, shown here for sake of later discussion.

In this paper we are interested in what happens when we vary the control
parameters in the Bose-Fermi coupling, $A,\Gamma ,\Lambda$, 
of Eq.~(\ref{Eq:MON}). This problem is
similar to (but more complex than) that of a bosonic system in an external
field already discussed years ago by Landau~\cite{landau} where one is
interested in what happens as one varies the strength of the field, and to
that of a fermion in a bath of harmonic oscillator bosons recently discussed
in condensed matter physics~\cite{lehur}.

\section{Classical analysis}
\label{Sec3}

Since the system we are considering here is that of an ensemble of bosons
coupled to a single fermion, it is convenient to analyze the situation in
terms of the motion of the single particle in the external field generated
by the bosons~\cite{iac1}. This is done by introducing a (number projected)
boson condensate and evaluating the expectation value of the Hamiltonian $H$
in the condensate~\cite{lev88,levshao89}. 
This produces the fermion single particle Hamiltonian in
the boson field, ${\cal H}$. 
Diagonalization of ${\cal H}$ gives the
single particle energies. The total energy is the sum of the single particle
energy and of the boson energy. Minimization of this total energy with
respect to the boson classical variables gives the equilibrium values
(classical order parameters). The behavior of the order parameter(s) as a
function of the control parameter(s) determines the nature and order of the
phase transition.

\subsection{Expectation value of H$_{B}$ in the boson condensate}
\label{subSec3.1}

We begin by considering the boson condensate with good particle 
number~$N$~\cite{diep,ginocchio,bohr} 
\begin{equation}
\left\vert N;\beta ,\gamma \right\rangle =\frac{1}{\sqrt{N!}}\left[
b_{c}^{\dag }(\beta ,\gamma )\right] ^{N}\left\vert 0\right\rangle
\label{Eq:cond}
\end{equation}
with 
\begin{equation}
b_{c}^{\dag }(\beta ,\gamma )=\frac{1}{\left( 1+\beta ^{2}\right) ^{1/2}}
[\, \beta \cos \gamma\, d_{0}^{\dag } 
+\frac{1}{\sqrt{2}}\beta \sin \gamma\,
( d_{2}^{\dag }+d_{-2}^{\dag }) +s^{\dag }\, ] ~.
\label{Eq:bc}
\end{equation}
The expectation value of the boson Hamiltonian $H_{B}$ of 
Eq.~(\ref{Eq:hBhFvBF}) in this condensate is given by~\cite{zamfir} 
\begin{eqnarray}
E_{B}(N;\beta ,\gamma ) &=&\left\langle N;\beta ,\gamma \left\vert
H_{B}\right\vert N;\beta ,\gamma \right\rangle   
\nonumber \\
&=&\varepsilon _{0}
N\left\{\left (\frac{\beta ^{2}}{1+\beta ^{2}}\right )
\left [
( 1-\xi ) -( \chi ^{2}+1) \frac{\xi }{4N}\right ] 
-\frac{5\xi }{4N(1+\beta ^{2})}\right.  
\nonumber \\
&&\qquad\quad\left. -\frac{\xi }{4(1+\beta ^{2})^{2}}\frac{N-1}{N}
\left[ 4\beta ^{2}-4\sqrt{\frac{2}{7}}\chi \beta ^{3}\cos 3\gamma 
+\frac{2}{7}\chi ^{2}\beta ^{4}\right] \right\} ~.\qquad
\label{Eq:EB}
\end{eqnarray}
The limit $N\rightarrow \infty $ of this expectation value is of interest
and it is 
\begin{eqnarray}
\bar{E}_{B}(\beta ,\gamma ) &=&
\lim_{N\rightarrow \infty }E_{B}(N;\beta,\gamma ) 
\nonumber \\
&=&\varepsilon _{0}N \left\{\left( 
\frac{\beta ^{2}}{1+\beta ^{2}}\right) \left(1-\xi \right)
\right.  
\nonumber \\
&&
\left.
\qquad\quad
-\frac{\xi }{4}\frac{1}{\left( 1+\beta ^{2}\right) ^{2}}\left[ 4\beta
^{2}-4\sqrt{\frac{2}{7}}\chi \beta ^{3}\cos 3\gamma +\frac{2}{7}\chi
^{2}\beta ^{4}\right] \right\} ~.
\label{Eq:EBlargeN}
\end{eqnarray}

\subsection{Expectation value of H$_{F}$ and V$_{BF}$ in the boson 
condensate}
\label{subSec3.2}

By integrating out the boson degrees of freedom, i.e. by taking the
expectation value of $H_{F}$ and $V_{BF}$ in the boson condensate, one
obtains the fermion Hamiltonian
\begin{eqnarray}
{\cal H}(N;\beta ,\gamma ) &=& E_{B}(N;\beta ,\gamma )  
\nonumber \\
&&+\sum_{m_{1},m_{2}}\left[\, \varepsilon _{j}\,
\delta_{m_{1},m_{2}}+g_{m_{1},m_{2}}(N;\beta ,\gamma )\, \right]  
\left( 
\frac{a_{j,m_{1}}^{\dag }a_{j,m_{2}}+a_{j,m_{2}}^{\dag}a_{j,m_{1}}}
{1+\delta _{m_{1},m_{2}}}\right) ~.
\quad
\label{Eq:Hdef}
\end{eqnarray}
The matrix $g_{m_{1},m_{2}}$ is a real, symmetric matrix, with explicit form
given by
\begin{equation}
g_{m_{1},m_{2}}(N;\beta ,\gamma ) = 
g_{m_{1},m_{2}}^{MON}(N;\beta ,\gamma)
+ g_{m_{1},m_{2}}^{QUAD}(N;\beta ,\gamma )
+ g_{m_{1},m_{2}}^{EXC}(N;\beta,\gamma )
\label{Eq:gmm}
\end{equation}
and
\begin{eqnarray}
g_{m_{1},m_{2}}^{MON} &=&NA\left( \frac{\beta ^{2}}{1+\beta ^{2}}\right)
\delta _{m_{1},m_{2}}\equiv NA\tilde{g}_{m_{1},m_{2}}^{MON} 
\label{Eq:gmMON}
\end{eqnarray}
\begin{eqnarray}
g_{m_{1},m_{2}}^{QUAD} &=&
N\Gamma \left( \frac{\beta }{1+\beta ^{2}}\right)(-)^{j+m_{2}}  
\nonumber \\
&&\times \left\{ \left[ 2\cos \gamma 
-\chi \sqrt{\frac{2}{7}}\beta \cos 2\gamma \right] 
\left\langle j,m_{1};j,-m_{2}\mid 2,0\right\rangle 
\delta_{m_{1},m_{2}}\right.  
\nonumber \\
&&\left. \qquad
+\left[ \sqrt{2}\sin \gamma +\chi \sqrt{\frac{1}{7}}\beta \sin 2\gamma 
\right] \left\langle j,m_{1};j,-m_{2}\mid 2,2\right\rangle \right\}  
\nonumber \\
&\equiv &-N\Gamma \tilde{g}_{m_{1},m_{2}}^{QUAD} 
\label{Eq:gmQUAD}
\end{eqnarray}
\begin{eqnarray}
g_{m_{1},m_{2}}^{EXC} &=&N\Lambda \left( \frac{\beta ^{2}}{1+\beta ^{2}}
\right) \left\{\left[\frac{X_{0}}{\sqrt{5(2j+1)}}\right.\right.  
\nonumber \\
&& \left.\left.
+(-)^{j+m_{2}}\left (\sqrt{\frac{2}{7}}X_{2}\cos 2\gamma \left\langle
j,m_{1};j,-m_{2}\mid 2,0\right\rangle\right.\right.\right.  
\nonumber \\
&&\left.\left.\left. 
\qquad\qquad\qquad
-\frac{1}{\sqrt{280}}X_{4}\left( 7+5\cos 2\gamma \right) 
\left\langle j,m_{1};j,-m_{2}\mid 4,0\right\rangle \right)\right]
\delta _{m_{1},m_{2}}\right.  
\nonumber \\
&&
\left.
+\left( -\right) ^{j+m_{2}} \left [-\sqrt{\frac{1}{7}}X_{2}\sin 2\gamma
\left\langle j,m_{1};j,-m_{2}\mid 2,2\right\rangle
\right.\right.  \nonumber \\
&&
\left.\left.
\qquad\qquad\qquad
-\sqrt{\frac{3}{28}}X_{4}\sin 2\gamma 
\left\langle j,m_{1};j,-m_{2}\mid 4,2\right\rangle 
\right.\right.  
\nonumber \\
&&
\left.\left.
\qquad\qquad\qquad
-\frac{1}{4}\left( 1-\cos 2\gamma \right) X_{4}\left\langle
j,m_{1};j,-m_{2}\mid 4,4\right\rangle \right]\right\}  
\nonumber \\
&
\equiv &N\Lambda \tilde{g}_{m_{1},m_{2}}^{EXC} ~.
\label{Eq:gmEXC}
\end{eqnarray}
Here $m_{1}\geq m_{2}$ and 
\begin{equation}
X_{L}=(2j+1)\left\{ 
\begin{array}{ccc}
j & 2 & j \\ 
L & j & 2
\end{array}
\right\} ~.
\label{Eq:XL}
\end{equation}
(The values for $m_{1}<m_{2}$ can be obtained by noting that the matrix is
symmetric. The symbol $\left\langle .. \mid ..\right\rangle $ 
in Eqs.~(\ref{Eq:gmQUAD}) and (\ref{Eq:gmEXC}) denotes 
Clebsch-Gordan coefficients and the curly bracket in~(\ref{Eq:XL}) is 
a~$6j$-symbol.) Also, in Eqs.(11)-(13) we have isolated the dependence 
on $N$ and on the parameters $A,\Gamma ,\Lambda $, from that on the 
intrinsic variables $\beta $ and $\gamma $ by introducing the matrices 
$\tilde{g}^{MON},\tilde{g}^{QUAD},\tilde{g}^{EXC}$. 
The analysis of this section becomes then
independent of the values of the couplings parameters and of $N$.

\subsection{Diagonalization of the \~{g}-matrix}
\label{subSec3.3}

The basis for the diagonalization of $\tilde{g}$ is the fermion
single-particle basis
\begin{equation}
\left\vert j,m\right\rangle =
a_{j,m}^{\dag }\left\vert 0\right\rangle 
\qquad\qquad m=\pm j,\pm (j-1),...,\pm \frac{1}{2} ~.
\label{Eq:aj}
\end{equation}
Because of the structure of $\tilde{g}$, which couples basis states
differing by $\pm 2$ in $m$, the diagonalization of $\tilde{g}$ splits into
two (doubly degenerate) pieces with $m=j,j-2,j-4,...,-(j-1)$ and similarly
with $m\rightarrow -m$. The dimension of the basis is thus $j+\frac{1}{2}.$
(In the case we discuss here, $j=11/2$, the matrices are $6\times 6$. When $
j=3/2$ the matrices are $2\times 2$ and the results of the diagonalization
can be obtained in explicit analytic form. When $j=1/2$, the matrix is $
1\times 1$ and only $\tilde{g}^{MON}$ and the $X_{0}$ term in $\tilde{g}
^{EXC}$ contribute trivially).

The diagonalization of the matrix $\tilde{g}_{m_{1},m_{2}}(\beta ,\gamma )$
yields the single particle eigenvalues 
\begin{equation}
\varepsilon _{i}(\beta ,\gamma ;\chi ) 
\qquad\qquad i=1,2,...,j+\frac{1}{2}
\label{Eq:epsiloni}
\end{equation}
and eigenfunctions
\begin{equation}
\left\vert \psi _{i}(\beta ,\gamma ;\chi )\right\rangle
=\sum_{m}c_{m}^{i}(\beta ,\gamma ;\chi )\left\vert j,m\right\rangle ~.
\label{Eq:efi}
\end{equation}
These are the single particle levels in the deformed $\beta $ and $\gamma $
field generated by the bosons. They depend on the boson parameter $\chi $
which distinguishes different types of boson fields ($\chi =0$, SO(6)
symmetry; $\chi =-\frac{\sqrt{7}}{2}$, SU(3) symmetry).

The contribution to the classical energy coming from the Bose-Fermi
interaction is the single particle energy
\begin{equation}
e_{i}(N;\beta ,\gamma ;\chi ;A,\Gamma ,\Lambda )=N\left[ A\varepsilon
_{i}^{MON}(\beta ,\gamma ;\chi )
-\Gamma \varepsilon _{i}^{QUAD}(\beta,\gamma ;\chi )
+\Lambda \varepsilon _{i}^{EXC}(\beta ,\gamma ;\chi )\right] ~.
\label{Eq:ei}
\end{equation}
This contribution is proportional to the number of bosons $N$ as it should
since the Bose-Fermi interaction, $V_{BF}$, is linear in the generators of
U(6).

\subsubsection{Results for the single particle eigenvalues}

(a) The monopole term. This term is diagonal with eigenvalues which do not
depend on the index $i$ and on the parameter $\chi $ 
\begin{equation}
\varepsilon _{i}^{MON}(\beta ,\gamma ;\chi )
=\left( \frac{\beta ^{2}}{1+\beta ^{2}}\right) ~.
\label{Eq:eiMON}
\end{equation}

\begin{figure}[t!]
\centering
\includegraphics[width=0.6\hsize]{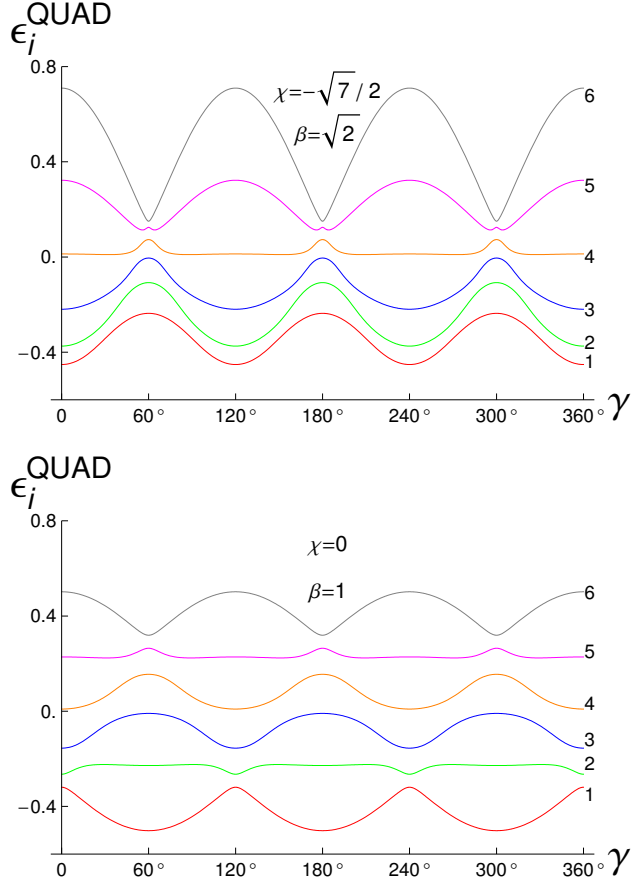}
\caption{Dependence on $\gamma $ of the eigenvalues 
$\varepsilon_{i}^{QUAD}\left( \beta ,\gamma ;\chi \right) $ 
($i=1,...,6)$ of the matrix 
$\tilde{g}_{m_{1},m_{2}}^{QUAD}$~(\ref{Eq:gmQUAD}) 
for $\chi =-\frac{\sqrt{7}}{2},\beta =\sqrt{2}$ (top part) and 
$\chi =0,\beta =1$ (bottom part), labelled in order of
increasing energy.}
\label{fig2}
\end{figure}

(b) The quadrupole term. The eigenvalues of the matrix 
$\tilde{g}_{m_{1},m_{2}}^{QUAD}$ must be obtained numerically. 
A~program has been
written in Mathematica to diagonalize the matrix. Using this program, we
have first investigated the symmetries of the eigenvalues. 
Fig.~\ref{fig2} shows the
eigenvalues $\varepsilon _{i}^{QUAD}(\beta,\gamma;\chi)$ 
as a function of $\gamma $ in the
domain $0\leq \gamma \leq 2\pi $ for $\chi =-\frac{\sqrt{7}}{2}$ and 
$\beta =\sqrt{2}$ (top part) and for $\chi =0$, $\beta =1$ (bottom part).
(We have chosen these values of $\beta $ because they are the equilibrium
values, $\beta _{e,B}$, of the boson energy functional, 
$\bar{E}_{B}(\beta,\gamma )$, Eq.~(\ref{Eq:EBlargeN}), 
for SU(3), $\chi =-\frac{\sqrt{7}}{2}$ and SO(6), $\chi =0$).
The various states are labelled by the index $i=1,...,6$ in order of
increasing energy. This figure shows the remarkable result that the single
particle energies generated by the quadrupole Bose-Fermi interaction are
periodic in $\gamma $ with period $\frac{2\pi }{3}$ and are symmetric under
reflections around $\frac{\pi }{3}$. This is in spite of the fact that 
$\tilde{g}$ contains terms 
$\cos \gamma ,\sin \gamma ,\cos 2\gamma ,\sin 2\gamma $.
We can therefore restrict our study to $0\leq \gamma \leq \frac{\pi }{3}$.
(The periodicity $\frac{2\pi }{3}$ can be proven by constructing the
characteristic polynomial that diagonalizes the matrix and showing that it
is a function of $\cos 3\gamma $). 
From Fig.~\ref{fig2} one can see that the 
$\gamma $ dependence of the single particle levels is different 
for $\chi =-\frac{\sqrt{7}}{2}$ than for $\chi =0$.

\begin{figure}[t!]
\centering
\includegraphics[width=0.6\hsize]{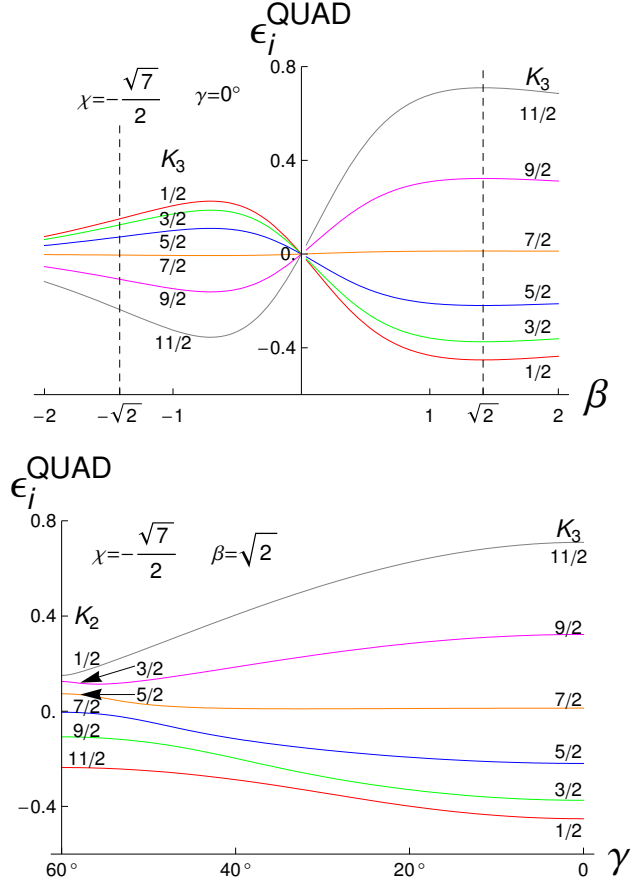}
\caption{Eigenvalues $\varepsilon _{i}^{QUAD}(\beta ,\gamma ;\chi )$ for a
particle with angular momentum $j=11/2$ in a quadrupole $\beta ,\gamma $
field with $\chi =-\frac{\sqrt{7}}{2}$ (prolate field) as a function of $
\beta $ for $\gamma =0^{\circ }$ (top part) and as a function of $\gamma $
for $\beta =\sqrt{2}$ (bottom part). The eigenvalues are labelled by the
projection of the angular momentum on the $\hat{3}$ axis at $\gamma =0^{0}$
and on the $\hat{2}$ axis at $\gamma =60^{\circ }$.}
\label{fig3}
\end{figure}

\begin{figure}[h!]
\centering
\includegraphics[width=0.6\hsize]{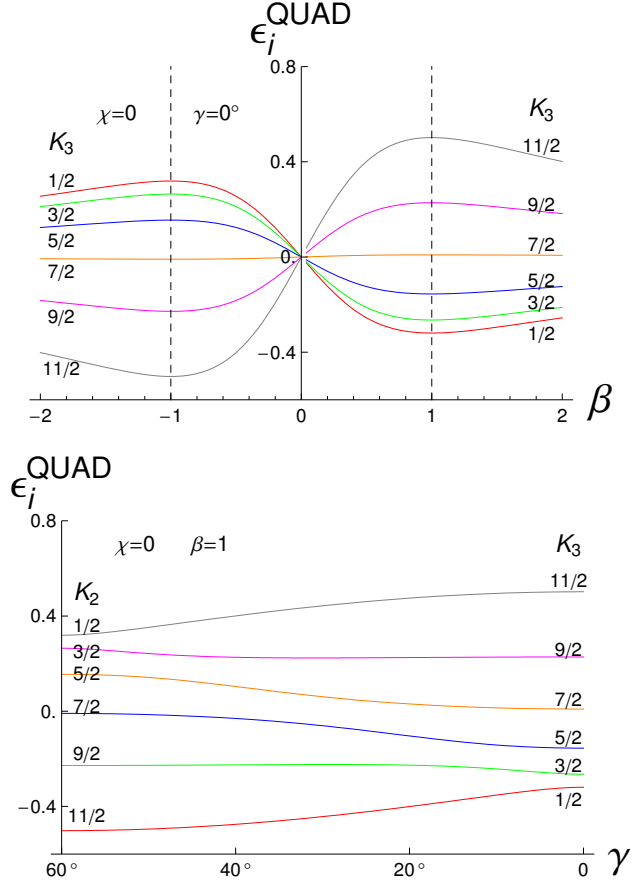}
\caption{Eigenvalues $\varepsilon _{i}^{QUAD}(\beta ,\gamma ;\chi )$ for a
particle with angular momentum $j=11/2$ in a quadrupole $\beta ,\gamma $
field with $\chi =0$ ($\gamma $-unstable field) as function of $\beta $ for $
\gamma =0^{\circ }$ (top part) and as a function of $\gamma $ for $\beta =1$
(bottom part). Labelling as in Fig.~\ref{fig3}.}
\label{fig4}
\end{figure}

In Figs.~\ref{fig3} and~\ref{fig4} 
we show the eigenvalues $\varepsilon _{i}^{QUAD}(\beta
,\gamma ;\chi )$ as a function of $\beta $ (at $\gamma =0^{\circ }$), in the
interval $-2\leq \beta \leq 2$ (top part), and as a function of $\gamma $ in
the interval $0^{\circ }\leq \gamma \leq 60^{\circ }$ at $\beta =\sqrt{2}$
for $\chi =-\frac{\sqrt{7}}{2}$ and at $\beta =1$ for $\chi =0$ (bottom
part). The bottom part is identical to Fig.~\ref{fig2} but restricted to 
$0^{\circ}\leq \gamma \leq 60^{\circ }$. 
We see that when $\chi =0$ we have an additional symmetry
\begin{equation}
\varepsilon _{i}^{QUAD}( \beta ,\gamma ;0) = 
-\varepsilon _{i}^{QUAD}(-\beta,\gamma ;0) ~.
\label{Eq:ei20}
\end{equation}
This symmetry is lost when $\chi \neq 0$. By combining the symmetries in $
\gamma $ with those in $\beta $ we find that the single particle energy
eigenvalues $\varepsilon _{i}^{QUAD}(\beta ,\gamma ;\chi )$ are invariant,
when $\chi =0$, under the transformation
\begin{equation}
\varepsilon _{i}^{QUAD}(-\beta ,\gamma =60^{\circ };0) =
\varepsilon _{i}^{QUAD}(\beta,\gamma =0^{\circ };0) ~.
\label{Eq:ei21}
\end{equation}
This symmetry is lost when $\chi \neq 0$. In order to recover it, we need to
simultaneously change the boson field from $\chi <0$ (prolate) to $\chi >0$
(oblate) when we go from $\beta >0$ to $\beta <0$. The single particle
energies have then the property 
\begin{equation}
\varepsilon _{i}^{QUAD}(-\beta ,\gamma =60^{\circ };-\chi ) = 
\varepsilon _{i}^{QUAD}(\beta,\gamma =0^{\circ };\chi ) ~.
\label{Eq:ei22}
\end{equation}
For the purpose of this article, where we are interested in the effect of a
single fermion on a bosonic system with a definite value of $\chi \leq 0$
(prolate boson field) it is sufficient to restrict all further analysis to
the domain $\beta \geq 0,$ $0^{\circ }\leq \gamma \leq 60^{\circ }$.

We also note that the matrix $\tilde{g}_{m_{1},m_{2}}^{QUAD}$ at $\gamma
=0^{\circ }$ is diagonal. Its eigenvalues are given in terms of the
projection $K_{3}$ of the angular momentum on the intrinsic axis $\hat{3}$.
The eigenvalues at $\gamma =60^{\circ }$ are instead given in terms of the
projection $K_{2}$ on the intrinsic axis $\hat{2}$. There is a continuous
change from $K_{3}$ to $K_{2}$ with the correspondence shown in 
Figs.~\ref{fig3} and~\ref{fig4}. 
The single particle levels in a $\beta ,\gamma $ field were studied years 
ago by Meyer-ter-Vehn~\cite{mayer} within the framework of the collective 
model. The $\gamma $ dependence of our results for $\chi =0$ is identical to
that of~\cite{mayer}. On the other side the $\beta $ dependence differs for
large $\beta $ due to the denominators $\frac{1}{1+\beta ^{2}}$ in our $
\tilde{g}$-matrix and to the fact that the variable $\beta $ of the
collective model, $\beta _{BM}$, is related to that of the Interacting Boson
Model, $\beta _{IBM}$, by a scale transformation, $\beta _{BM}=c\beta _{IBM}$
with $c\approx 0.2$ in the mass $A=150$ region~\cite{iac111}. We also note
that, if the variable $\gamma $ is frozen to $0^{\circ }$, one obtains the
single particle levels in a deformed $\beta $ field with axial symmetry.
Those were studied years ago by Nilsson within the framework of the
collective model~\cite{nilsson}. Our results for $\chi =0$ are similar to
those of the Nilsson model except for the denominators 
$\frac{1}{1+\beta ^{2}}$.

\begin{figure}[t!]
\centering
\includegraphics[width=0.6\hsize]{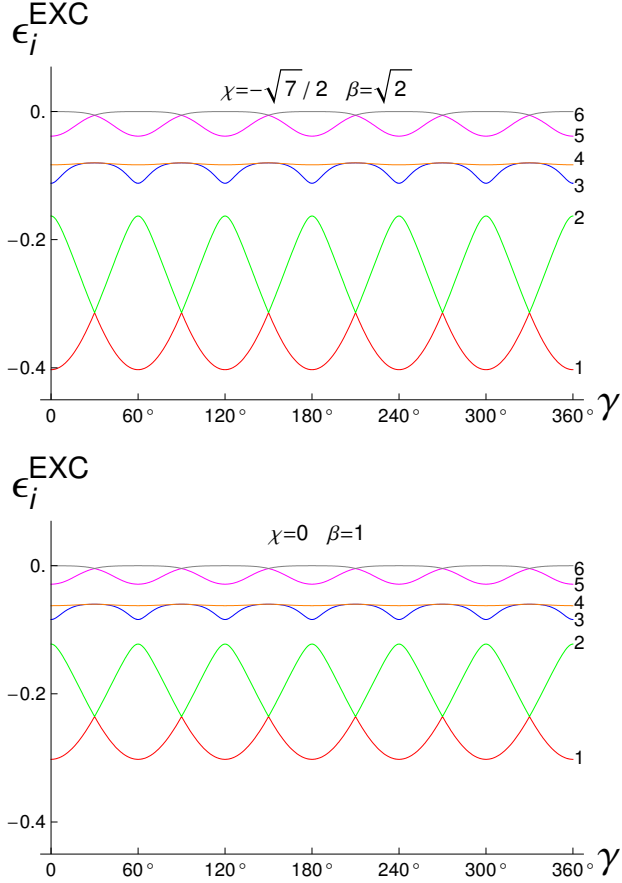}
\caption{Dependence on $\gamma $ of the eigenvalues 
$\varepsilon_{i}^{EXC}\left( \beta ,\gamma ;\chi \right)$ 
($i=1,...,6$) of the matrix $\tilde{g}_{m_{1},m_{2}}^{EXC}$ for 
$\chi =-\frac{\sqrt{7}}{2},\beta =\sqrt{2}$ (top part) 
and $\chi =0,\beta =1$ (bottom part), labelled in order of
increasing energy.}
\label{fig5}
\end{figure}

(c) The exchange term. The eigenvalues of the matrix 
$\tilde{g}_{m_{1},m_{2}}^{EXC}$ are also obtained numerically. 
We first investigate the symmetries of the eigenvalues. 
Fig.~\ref{fig5} shows the eigenvalues 
$\varepsilon_{i}^{EXC}(\beta,\gamma;\chi)$ 
as a function of $\gamma $ in the domain $0\leq \gamma \leq 2\pi 
$ for $\chi =-\frac{\sqrt{7}}{2},\beta =\sqrt{2}$ and 
$\chi =0,\beta =1$. 
This figure shows that the contributions to the single particle energies
coming from the exchange interaction are periodic in $\gamma $ with period 
$\frac{\pi }{3}$, that is half of the period of the quadrupole interaction,
and are symmetric under reflection around $\frac{\pi }{6}$, i.e. they are a
function of $\cos 6\gamma $. Furthermore, the exchange term is independent 
of $\chi$ and is an even function of $\beta$.  

\begin{figure}[t!]
\centering
\includegraphics[width=0.6\hsize]{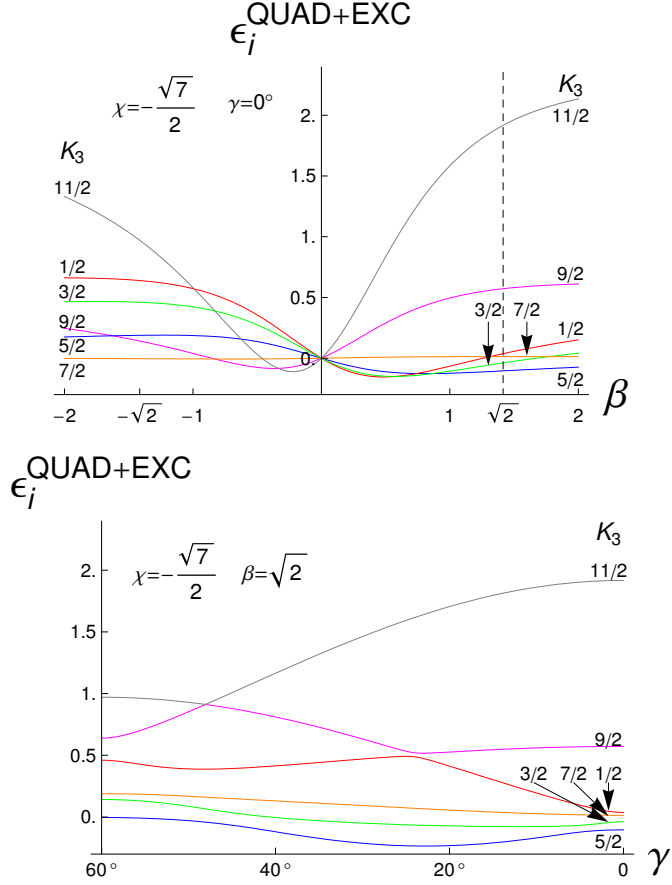}
\caption{Eigenvalues $\varepsilon _{i}^{QUAD+EXC}(\beta ,\gamma ;\chi )$ 
for a particle with $j=11/2$ in an external 
$\beta ,\gamma ;\chi =-\frac{\sqrt{7}}{2}$ field with both quadrupole 
and exchange interaction with strengths 
$\Lambda /\Gamma =3$. Labelling as in Fig.~\ref{fig3}. 
Figure constructed by using $\Gamma =-0.0958,\Lambda =-0.287$.}
\label{fig6}
\end{figure}

\begin{figure}[t!]
\centering
\includegraphics[width=0.6\hsize]{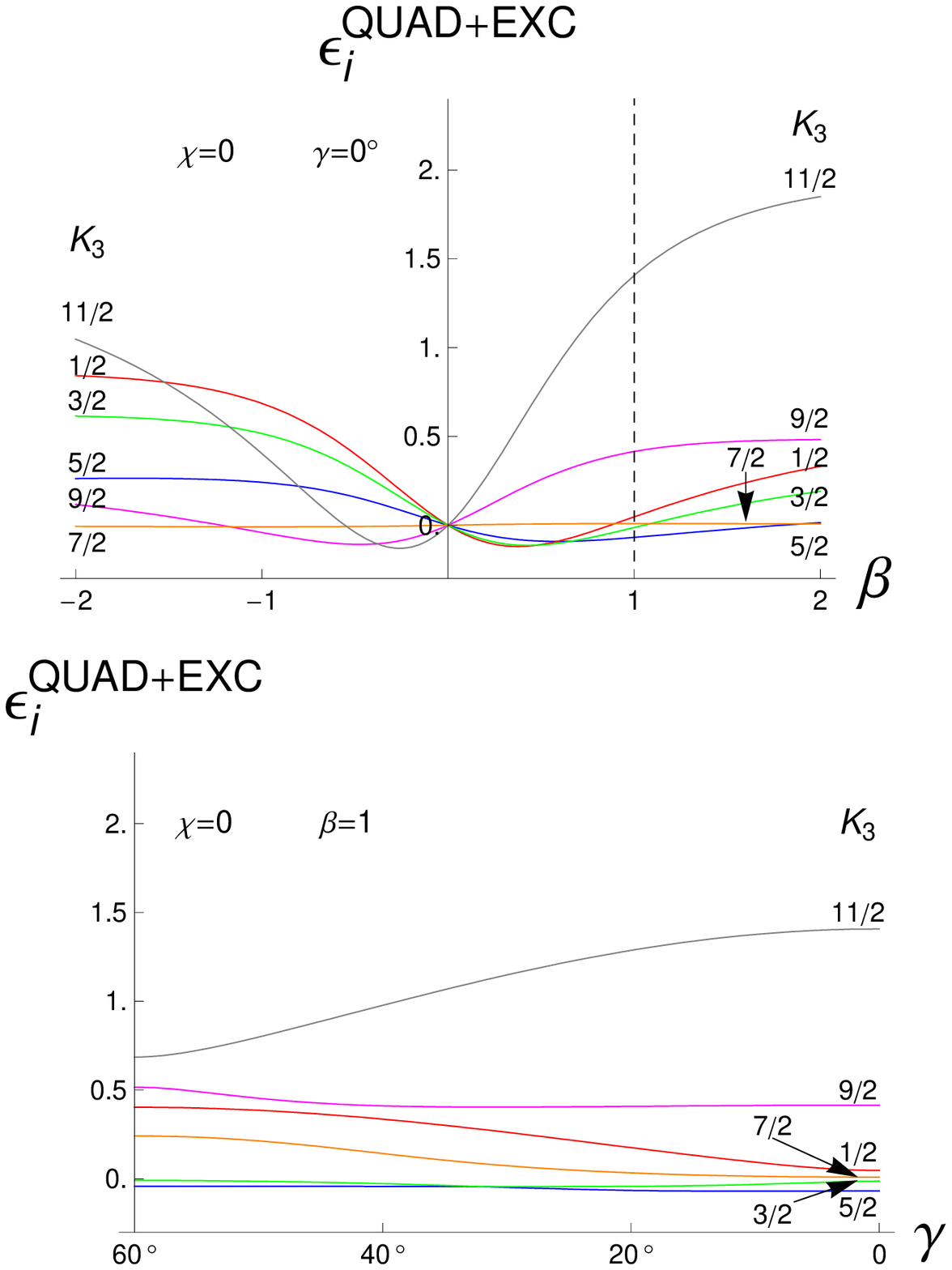}
\caption{Same as Fig.~\ref{fig6} but with $\chi =0$.}
\label{fig7}
\end{figure}

(d) Combination of quadrupole and exchange. We consider here the 
eigenvalues,\\
$\varepsilon _{i}^{QUAD+EXC}(\beta,\gamma ;\chi)$, 
of the matrix 
\begin{equation}
\tilde{g}_{m_{1},m_{2}}^{QUAD+EXC}(\beta ,\gamma ;\chi )
= -\tilde{g}_{m_{1},m_{2}}^{QUAD}(\beta ,\gamma ;\chi )
+\left( \frac{\Lambda }{\Gamma }\right) 
\tilde{g}_{m_{1},m_{2}}^{EXC}(\beta ,\gamma ;\chi )
\end{equation}
and, for sake of display, we take 
$\left( \frac{\Lambda }{\Gamma }\right) =3$. 
The eigenvalues of this matrix are shown in Figs.~\ref{fig6} 
and~\ref{fig7}.
We see from these figures that the ordering of the single-particle
eigenvalues at a given $\beta ,\gamma $ when both quadrupole and exchange
interactions are present is rather complex. Our results are similar to those
of the Nilsson model plus BCS~\cite{iac1}. The similarity is no longer so
obvious as in the case in which the coupling is purely quadrupole, but it
can be seen by solving the BCS equations in the deformed $\beta ,\gamma $
field as done by Meyer-ter-Vehn~\cite{mayer}.

The formalism is flexible enough that one can also investigate particle-hole
conjugation. Within the framework of IBFM, particle-hole conjugation is the
transformation $\Gamma \rightarrow -\Gamma ,\Lambda \rightarrow \Lambda $
(see the following Eq.~(\ref{Eq:IBFMcalcparam})). 
When $\Lambda =0$, the effect of this
transformation is to reverse the order of the single particle levels 
in the previous Figs.~\ref{fig3} and~\ref{fig4}, 
as one can simply see from the form of $g^{QUAD}(\beta ,\gamma ;\chi )$ 
in Eq.~(\ref{Eq:gmQUAD}). 
When $\Lambda \neq 0$ the transformation leads to a different ordering 
of the single particle levels. We do not dwell
on this point further, but note instead that a special circumstance occurs
when the variable $\gamma $ is frozen to $0^{\circ }$ 
(prolate axial symmetry). 
As remarked above, the quadrupole Bose-Fermi interaction is diagonal at 
$\gamma =0^{\circ }$. It turns out that also the exchange interaction is
diagonal. The eigenvalues of the $g$-matrix for the combined quadrupole plus
exchange interaction are given by~\cite{lev88} 
\begin{eqnarray}
\lambda_{K}(\beta ) &=& -N\Gamma \left\{ \left( \frac{\beta }{1+\beta^{2}}
\right) \sqrt{5}\left (\, 2-\beta \chi \sqrt{\frac{2}{7}}\,\right )
P_{j}[3K^{2}-j(j+1)]\right\}   
\nonumber \\
&&-N\Lambda \left\{ \left( \frac{\beta ^{2}}{1+\beta ^{2}}\right)
(2j+1)P_{j}^{2}[3K^{2}-j(j+1))]^{2}\right\} 
\label{Eq:lamK}
\end{eqnarray}
where
\begin{equation}
P_{j}=\left[ (2j-1)j(2j+1)(j+1)(2j+3)\right] ^{-1/2} ~.
\label{Eq:Pj}
\end{equation}
The eigenvalues can be labelled by the projection of the angular momentum on
the intrinsic axis $\hat{3}$, $K_{3}\equiv K=\frac{1}{2},\frac{3}{2},...,j$
and they are doubly degenerate.

The eigenvalues of the $g$-matrix for the combined quadrupole plus
exchange interaction are also diagonal at $\gamma=60^{\circ}$ 
(oblate axial symmetry), and are given by
\begin{eqnarray}
\omega_{K_2}(\beta ) &=& -N\Gamma \left\{ \left( 
\frac{\beta }{1+\beta^{2}}\right) 
\sqrt{5}\left (\, -2 -\beta \chi \sqrt{\frac{2}{7}}\,\right )
P_{j}[3K_{2}^{2}-j(j+1)]\right\}   
\nonumber \\
&&-N\Lambda \left\{ \left( \frac{\beta ^{2}}{1+\beta ^{2}}\right)
(2j+1)P_{j}^{2}[3K_{2}^{2}-j(j+1))]^{2}\right\} ~, 
\label{Eq:omegaK}
\end{eqnarray}
where $K_2$ is the projection of the angular momentum on the intrinsic 
axis $\hat{2}$. They satisfy the relation
\begin{equation}
\omega_{K_2}(\beta) = \lambda_{K_3\to K_2}(-\beta) ~.
\label{Eq:lamomeg}
\end{equation}

\subsection{Equilibrium values: classical order parameters}
\label{subSec3.4}

Having determined the single particle energies, $e_{i}$, 
Eq.~(\ref{Eq:ei}), we can construct the total energy surface
\begin{equation}
E_{i}(N;\beta ,\gamma ;\xi ,\chi ;A,\Gamma ,\Lambda )
= E_{B}(N;\beta ,\gamma;\xi ,\chi )+\varepsilon _{j} + 
e_{i}(N;\beta ,\gamma ;\chi ;A,\Gamma ,\Lambda) ~, 
\label{Eq:Ei}
\end{equation}
where $E_B$, Eq.~(\ref{Eq:EB}), and hence $E_i$ depend also on the 
scale parameter $\varepsilon_{0}$. 
In the traditional approach to odd-even nuclei, the equilibrium values 
$\beta _{e},\gamma _{e}$ are obtained by minimizing the boson energy $E_{B}$
(or core energy in the collective model). This provides a static deformation
and the single particle energies are evaluated in this static deformation.
This is a good approximation when the deformation is large. In the
transitional region, this is no longer the case. A better approximation is
to minimize the total energy surface including the contribution of the
fermions (the odd-particle). Although the fermion contribution is of order 
$1/N$, it will be shown that it has a dramatic effect on the phase
transition, modifying the location of the critical point.

\begin{figure}[t!]
\centering
\includegraphics[scale=0.5]{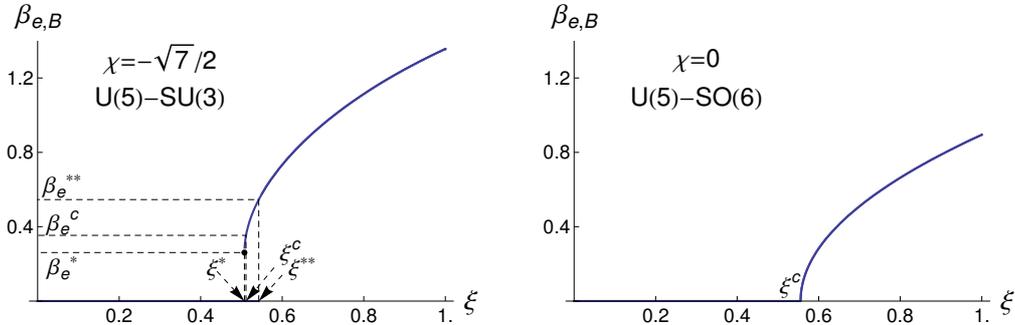}
\caption{Equilibrium values, $\beta _{e,B}$, as a function of the control
parameter, $\xi $, of the bosonic part of the energy functional, $E_{B}$, 
for the transition U(5)-SU(3), $\chi =-\frac{\sqrt{7}}{2}$, left panel, and
U(5)-SO(6), $\chi =0$, right panel. The spinodal, critical and antispinodal
points are denoted by $\xi ^{\ast },\xi ^{c},\xi ^{\ast \ast }$,
respectively. Figure constructed for $N=10$ bosons.}
\label{fig8}
\end{figure}

The equilibrium values $\beta _{e}$ and $\gamma _{e}$ (the classical order
parameters) for the combined system are obtained by minimizing $E_{i}$ with
respect to $\beta $ and $\gamma $, i.e. imposing the conditions
\begin{equation}
\frac{\partial E_{i}}{\partial \beta }=0 \;\; ,\;\;
\frac{\partial E_{i}}{\partial \gamma }=0 ~.
\label{Eq:partialEi}
\end{equation}
Minimization of $E_{i}$ is, in general, no simple matter. The boson part, 
$E_{B}$, has a minimum at $\beta _{e,B}(\xi )$ and at 
$\gamma _{e,B}=0^{\circ}$, for any value of $\xi $ and $\chi<0$, 
except for $\chi =0$ where there is
no minimum in $\gamma $ ($\gamma $-unstable situation). The boson
equilibrium values $\beta _{e,B}(\xi )$ have been calculated by several
authors~\cite{zamfir} and are shown in Fig.~\ref{fig8} 
as a function of the control parameter $\xi $ for two values of 
$\chi =-\frac{\sqrt{7}}{2}$ and $\chi =0$, 
for later comparison with those in odd-even nuclei (Bose-Fermi system).

The single particle eigenvalues $\varepsilon _{i}$ have a rather complex
structure (Figs.~\ref{fig3} and~\ref{fig4}), 
especially when there is an exchange interaction
(Figs.~\ref{fig6} and~\ref{fig7}). 
When doing the minimization, an important question is on
what space to minimize. In view of the periodicity in $\gamma$, it is
sufficient to minimize $E_{i}$ in the sextant 
$0^{\circ }\leq \gamma \leq 60^{\circ }$. 
Also, since the purpose of this article is to understand the
extent to which the presence of an odd fermion modifies the phase
transition, we seek solutions in the neighbourhood of the boson equilibrium
values and thus we do a constrained minimization with $\beta \geq 0$, in
practice $0\leq \beta \leq 2$.

To fully investigate QPTs in IBFM, one should do a study as a function of
all control parameters $\xi ,\chi $ (bosons) and $A,\Gamma ,\Lambda $
(Bose-Fermi). In this paper we study two phase transitions: (i) U(5) to
SU(3), spherical to axially deformed, $\chi =-\frac{\sqrt{7}}{2}$ and~(ii)
U(5) to SO(6), spherical to $\gamma $-unstable, $\chi =0$, both induced 
by a change in $0\leq \xi \leq 1$. These phase transitions have been 
investigated extensively in the purely bosonic case and are of practical 
importance. We also set $A=0$, since the monopole term only renormalizes 
the $\hat{n}_{d}$ term in the boson Hamiltonian, $H_{B}$, and its effects 
can be easily seen. We are left with two additional control parameters, 
$\Gamma $ and $\Lambda $. In order to isolate the dependence on the 
fermion angular momentum~$j$, we write
\begin{eqnarray}
\Gamma  &=&\Gamma_{0}\,Q_{jj}  
\nonumber \\
Q_{jj} &=&\left\langle j\parallel Y^{(2)}\parallel j\right\rangle = 
\sqrt{\frac{5}{4\pi}}\sqrt{2j+1}
\left\langle j,1/2;2,0\mid j,1/2\right\rangle
\label{Eq:Qjj}
\end{eqnarray}
where the double-bar in Eq.~(\ref{Eq:Qjj}) denotes a reduced matrix element. 
To further restrict the study, we take $\Gamma _{0}$ proportional to $\xi$
\begin{equation}
\Gamma _{0}=-2\varepsilon _{0}\frac{\xi }{4N} ~.
\label{Eq:Gam0}
\end{equation}
The entire study presented here is thus in terms of a single control
parameter $\xi $, plus the value of $\Lambda $ which, for convenience, can
also be written as 
\begin{equation}
\Lambda =\lambda \varepsilon _{0}
\label{Eq:Lambda}
\end{equation}
making all terms in $H$ then proportional to the scale factor 
$\varepsilon_{0}$. The parameterization of Eq.~(\ref{Eq:Gam0}), 
for the quadrupole term, was suggested by Alonso 
\textit{et al.}~\cite{alonso1} and it has the advantage that the combined 
Hamiltonian can be rewritten, apart from an overall constant, as
\begin{eqnarray}
H &=& 
\varepsilon _{0}\left\{\, 
\left( 1-\xi \right) \left( \hat{n}_{d}+\hat{n}
_{j}\right) -\frac{\xi }{4N}\left( \hat{Q}^{\chi }
+\hat{q}_{F}\right) \cdot \left( \hat{Q}^{\chi }+\hat{q}_{F}\right) \right. 
\nonumber\\
&& \left. 
\qquad\;
+ \lambda\, \sqrt{2j+1}:[ ( d^{\dag }\times \tilde{a}_{j})^{(j)}
\times ( \tilde{d}\times a_{j}^{\dag })^{(j)}]^{(0)}:\, \right\} ~,
\label{Eq:Hmodel}
\end{eqnarray}
where
\begin{eqnarray}
\hat{q}_{F} &=& Q_{jj}\,\hat{q}_{j} ~,
\label{Eq:qF}
\end{eqnarray}
and $\hat{q}_j=  (a_{j}^{\dagger}\times \tilde{a}_{j})^{(2)}$. 
If one multiplies $Q_{jj}$ by $-\sqrt{\pi}$, the corresponding 
Hamiltonian for $j=3/2$ and $\chi =\lambda=0$, has 
a Bose-Fermi symmetry SO(6) $\otimes$ SU(4) $\supset$ Spin(6)~\cite{iac1}, 
{\it i.e.}, the Hamiltonian is
invariant under interchange of the boson operators 
$\hat{n}_{d},\,\hat{Q}^{\chi }$ with the fermion operators 
$\hat{n}_{j},\,\hat{q}_j$.

\subsubsection{Results for $\Lambda =0$}

\begin{figure}[t!]
\centering
\includegraphics[width=0.7\hsize]{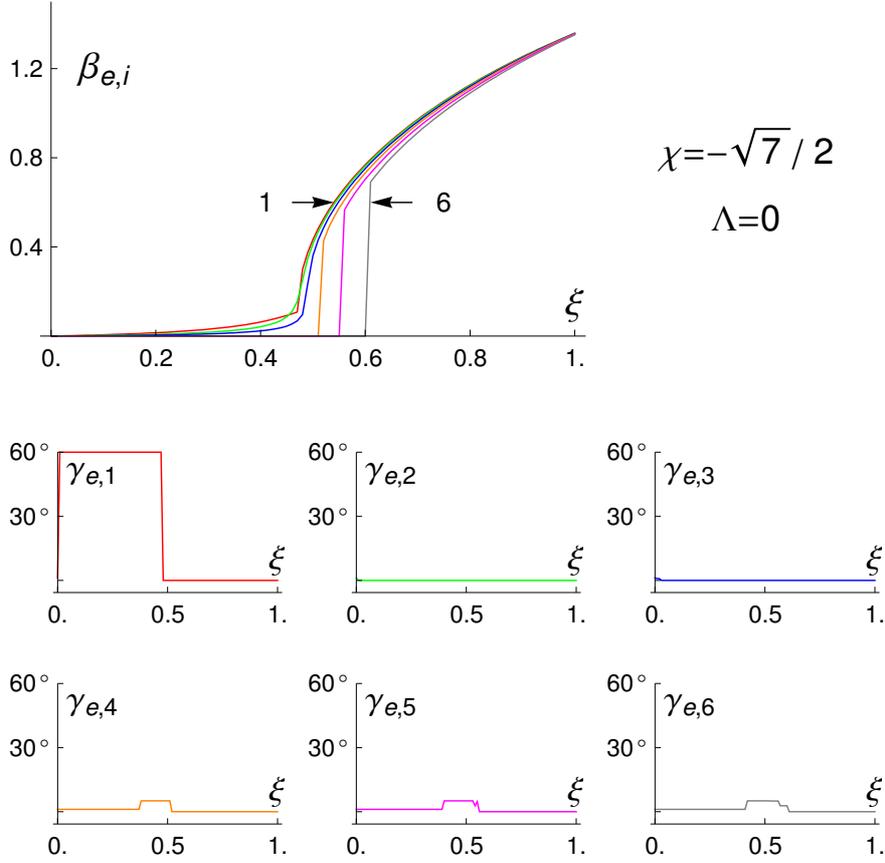}
\caption{Equilibrium values, $\beta _{e,i}$ (top part) and $\gamma _{e,i}$
(bottom part) as a function of the control parameter $\xi $ in the
U(5)-SU(3) transition ($\chi =-\frac{\sqrt{7}}{2}$), $\Lambda =0$ 
and $N=10$. States are labelled by the index $i=1,...,6$.}
\label{fig9}
\end{figure}

\begin{figure}[t!]
\centering
\includegraphics[width=0.7\hsize]{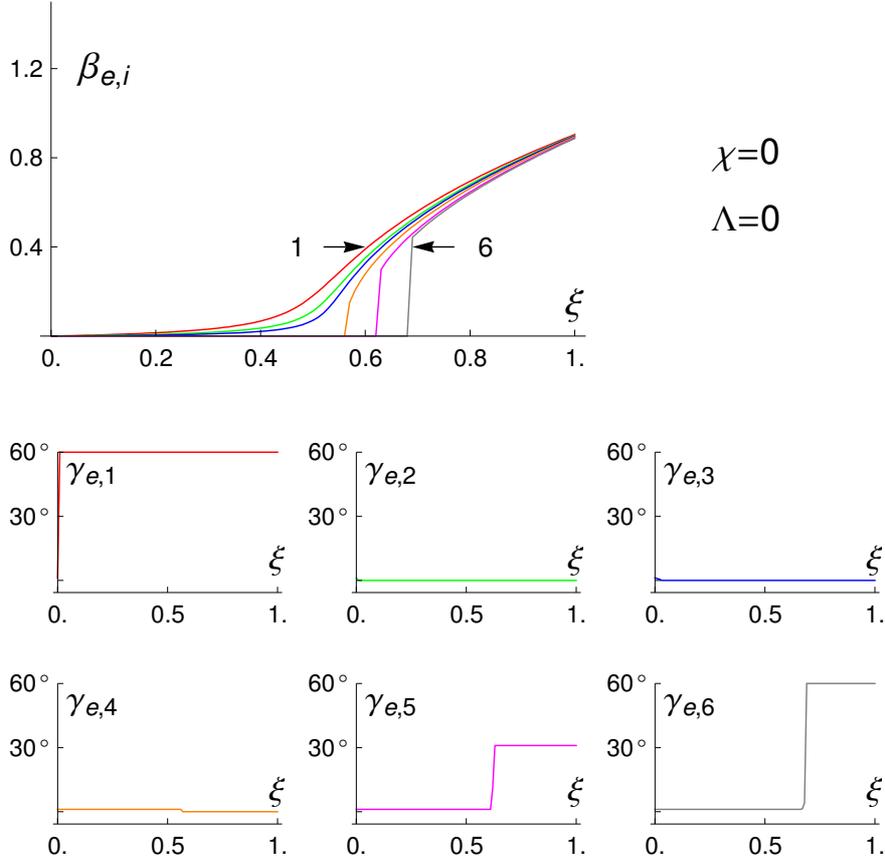}
\caption{Equilibrium values, $\beta _{e,i}$ (top part) and $\gamma _{e,i}$
(bottom part) as a function of the control parameter $\xi $ in the
U(5)-SO(6) transition ($\chi =0$), $\Lambda =0$ and $N=10$. 
Labelling as in Fig.~\ref{fig9}.}
\label{fig10}
\end{figure}

The equilibrium values for the U(5)-SU(3) transition are shown in 
Fig.~\ref{fig9} 
and, for the U(5)-SO(6) transition, in Fig.~\ref{fig10}.
From these figures one can see the effects of the odd particle on the phase
transition. In the U(5)-SU(3) case, the phase transition is washed out for
states~1,2,3 and enhanced for states~4,5,6. The critical point is
approximately at the same location as for the purely bosonic case 
($\xi_{B}^{c}\sim$0.51) for state~4 ($\xi _{4}^{c}\sim $0.50), but it 
is moved to larger values for state~5 ($\xi _{5}^{c}\sim $0.53) and state~6 
($\xi_{6}^{c}\sim $0.58). The values of $\gamma _{e}$ below and around the
critical point are no longer zero. After the critical point, all states
become those of a single particle in an axially deformed field ($\gamma
_{e}=0^{\circ }$) with equal deformation $\beta _{e,i}=\beta _{e,B}$. States
in this region can be labelled by the projection 
$K=1/2,\, 3/2,\, 5/2,\, 7/2,\, 9/2,\, 11/2$ of the angular momentum 
on the intrinsic $\hat{3}$ axis, corresponding to 
states~$i=1,...,6$. A~different
situation occurs for the U(5)-SO(6) transition. Although the equilibrium
values, $\beta _{e,i}$, follow the same behavior as for the U(5)-SU(3)
transition, the equilibrium values $\gamma _{e,i}$ do not. States 1 and 6
are oblate ($\gamma _{e}=60^{\circ }$), states 2,3,4 are prolate 
($\gamma_{e}=0^{\circ }$) and state 5 is triaxial 
($\gamma _{e}=30^{\circ }$). For
this phase transition, the equilibrium value of $\gamma $ is dictated by
that of the single particle states, $e_{i}$, since the boson part is $\gamma 
$ independent. The effect of the odd-particle is more dramatic here than in
the case of the U(5)-SU(3) transition.

\subsubsection{Results for $\Lambda \neq 0$}

\begin{figure}[t!]
\centering
\includegraphics[width=0.7\hsize]{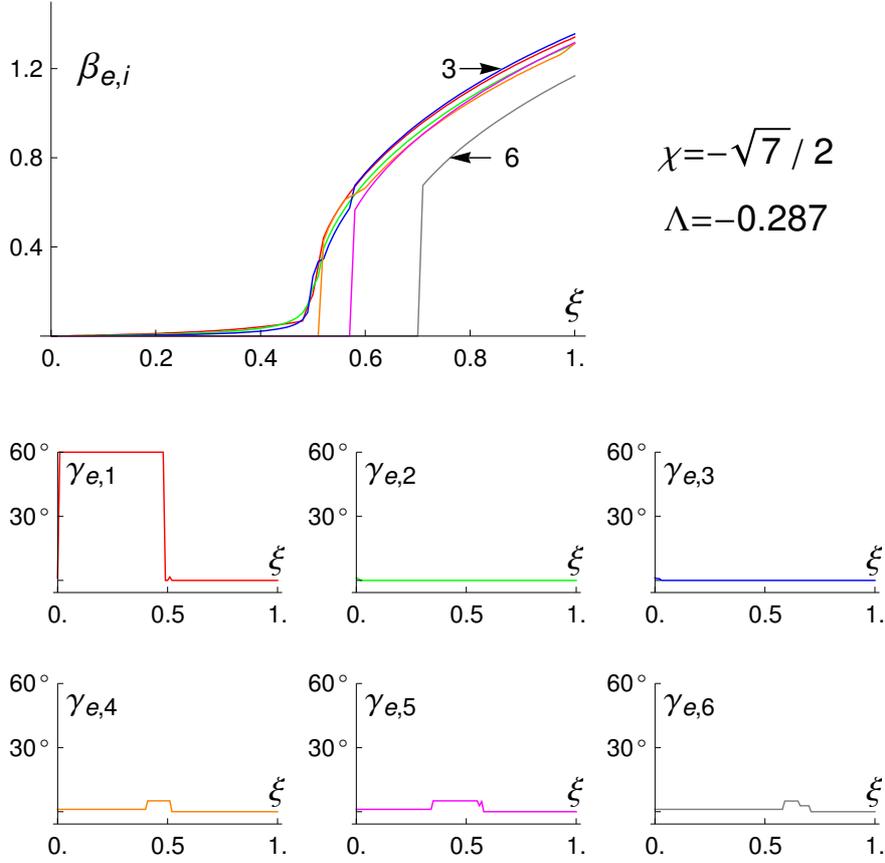}
\caption{Same as Fig.~\ref{fig9} but with $\Lambda =-0.287$.}
\label{fig11}
\end{figure}

\begin{figure}[h!]
\centering
\includegraphics[width=0.7\hsize]{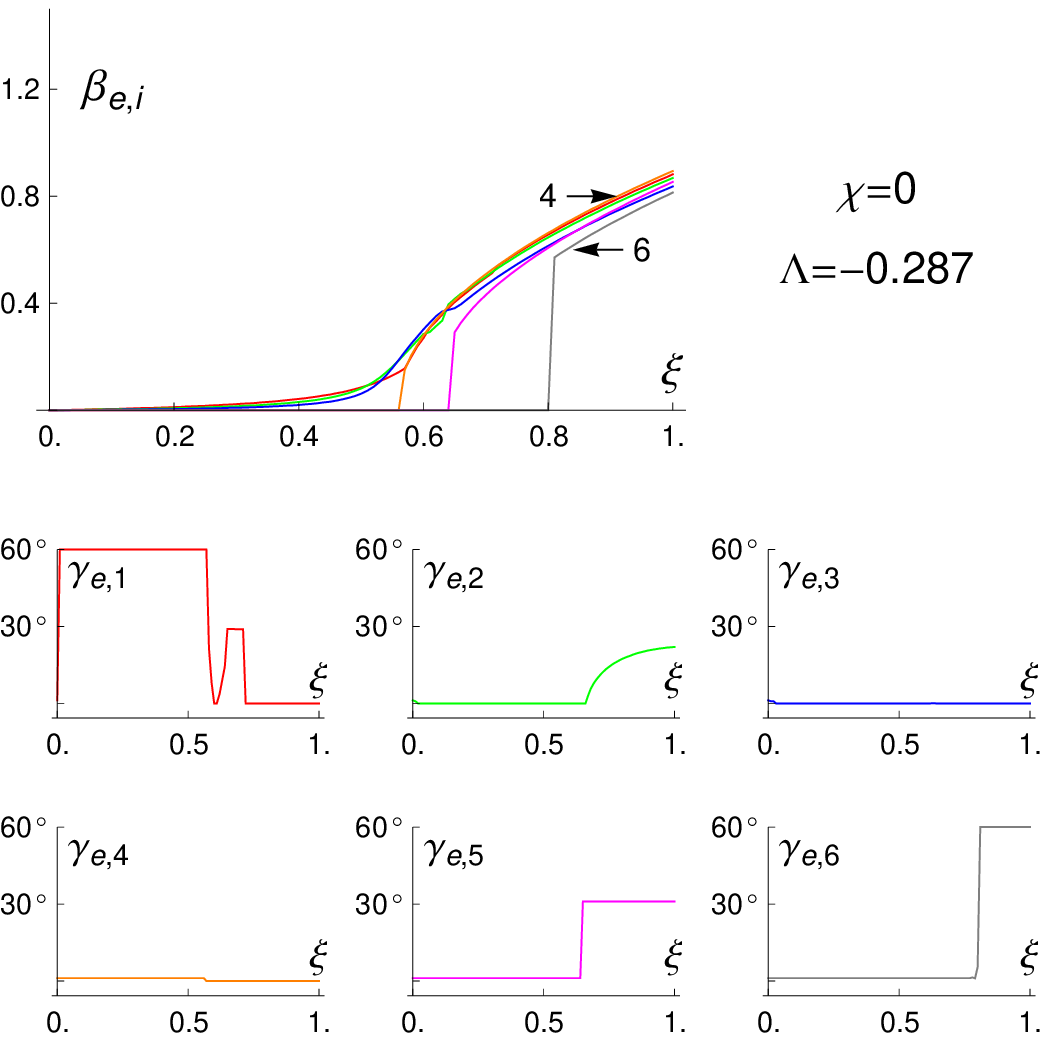}
\caption{Same as Fig.~\ref{fig10} but with $\Lambda =-0.287$.}
\label{fig12}
\end{figure}

The equilibrium values when $\Lambda =-0.287$ for the U(5)-SU(3) transition
are shown in Fig.~\ref{fig11}
and for the U(5)-SO(6) transition in Fig.~\ref{fig12}.
We see that the modifications induced by the presence of a fermion to the
phase transition are more dramatic when the exchange interaction is added, 
especially in the location of the critical point $\xi _{i}^{c}$. 
The state~6 moves considerably to the right of $\xi _{B}^{c}$. 
In summary, QPTs in the
presence of an odd fermion have a different behavior as a function of the
control parameter $\xi $ than in the purely bosonic system. The odd fermion
acts as a catalyst for some states and as a retardative for others.

\subsection{Single particle energies as a function of the control parameter 
$\protect\xi $}
\label{subSec3.5}

Once the values of $\beta_{e,i}$ and $,\gamma_{e,i}$ have been obtained, 
one can evaluate the single particle energies 
$e_{i}(N;\beta _{e,i},\gamma _{e,i})$ 
as a function of $\xi$. Throughout this section we set the scale parameter 
$\varepsilon_{0}=1$ and $N=10$.

\subsubsection{Results for $\Lambda =0$}

\begin{figure}[t!]
\centering
\includegraphics[scale=0.6]{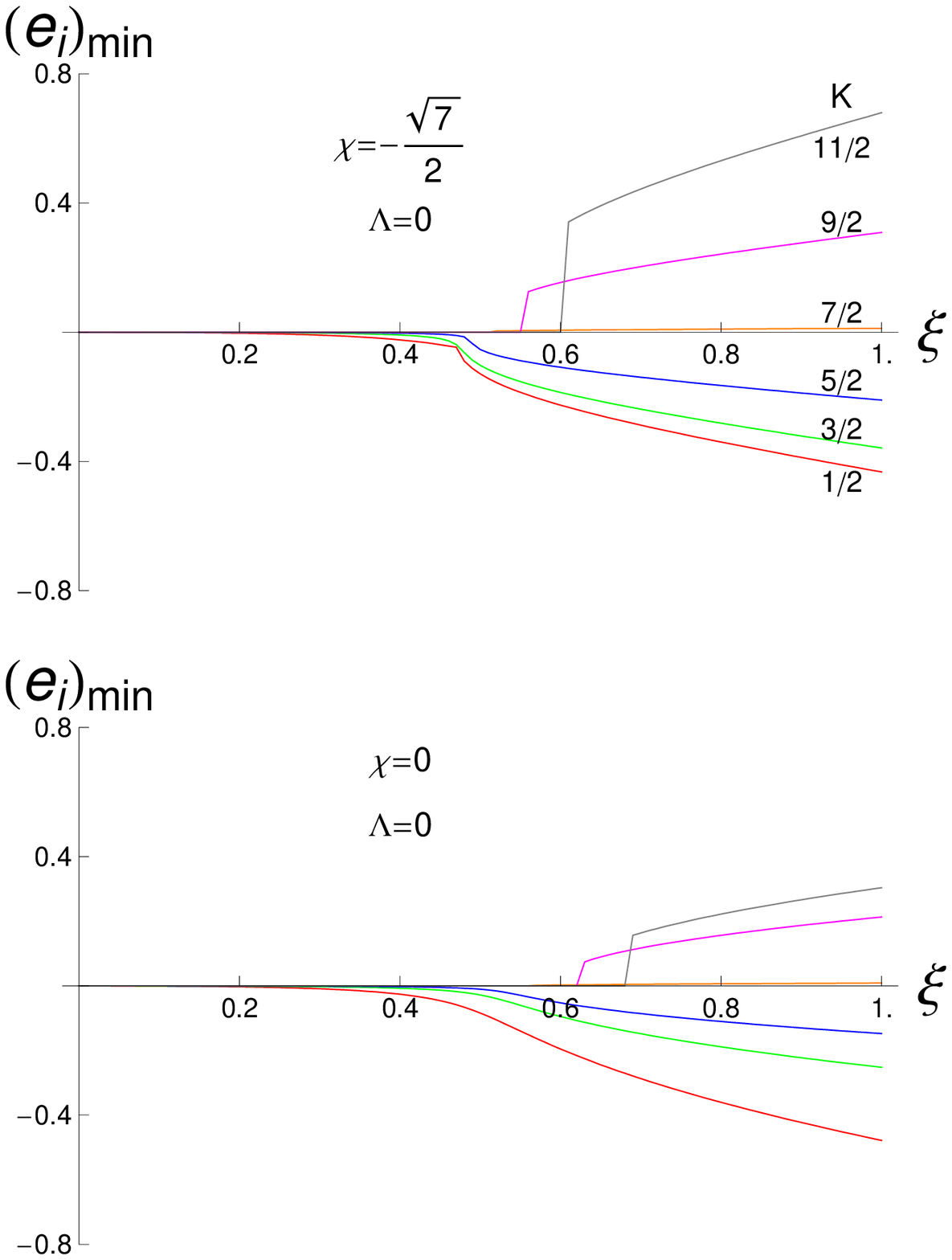}
\caption{Minimum single particle energies, 
$\left( e_{i}\right) _{\min}=
-N\Gamma \left( \varepsilon _{i}\right) _{\min }$ as a function of the
control parameter, $\xi $, in the U(5)-SU(3) transition 
($\chi =-\frac{\sqrt{7}}{2}$) (top part) and in the U(5)-SO(6) 
transition ($\chi =0$) (bottom part), both with $\Lambda =0$.}
\label{fig13}
\end{figure}

Results for $\chi =-\frac{\sqrt{7}}{2}$, $\chi =0$ and $\Lambda =0$ are
shown in Fig.~\ref{fig13}. 
The discontinuities reflect the phase transition.
\begin{figure}[t!]
\centering
\includegraphics[scale=0.6]{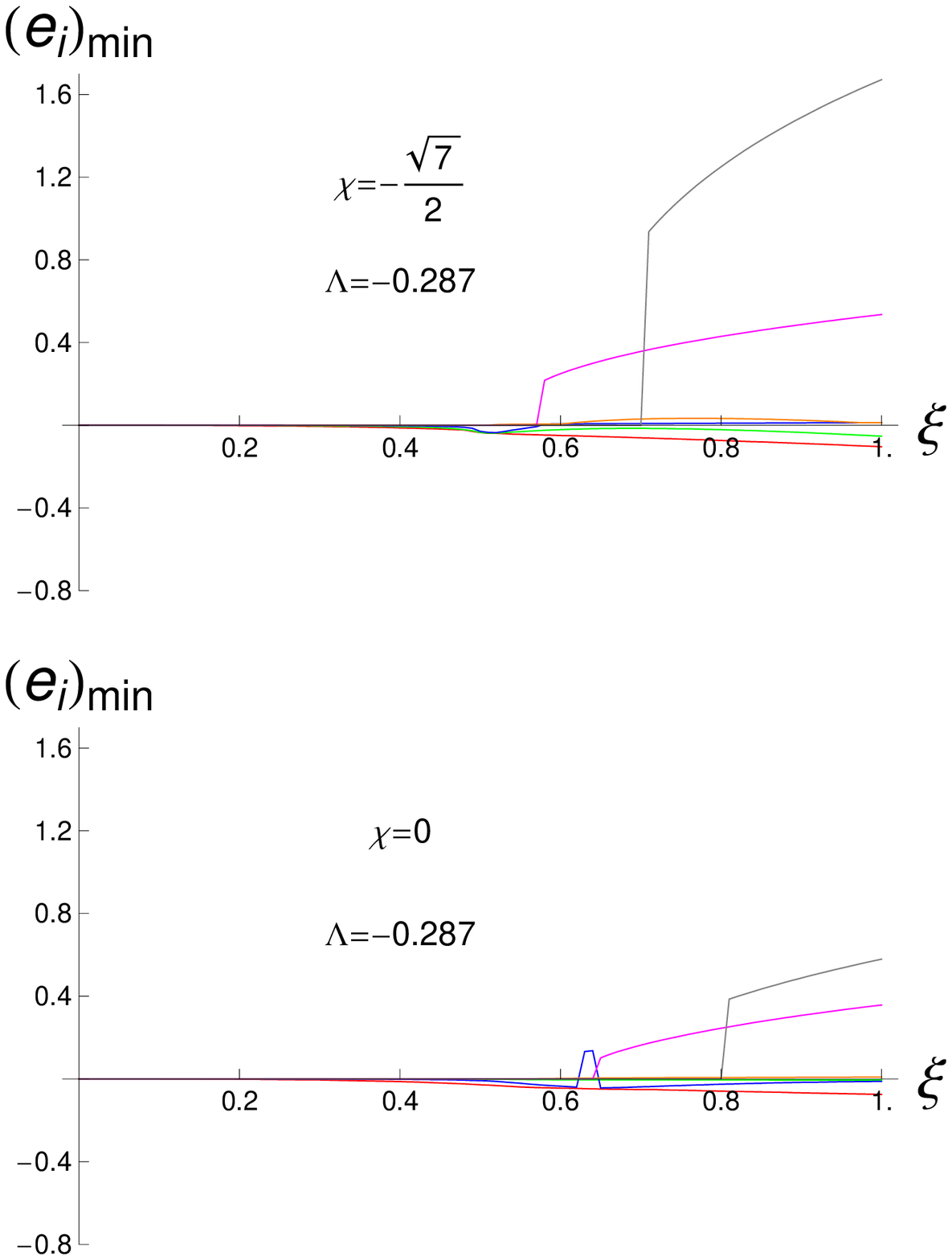}
\caption{Same as Fig.~\ref{fig13} but with $\Lambda =-0.287$.}
\label{fig14}
\end{figure}
\begin{figure}[t!]
\centering
\includegraphics[scale=0.6]{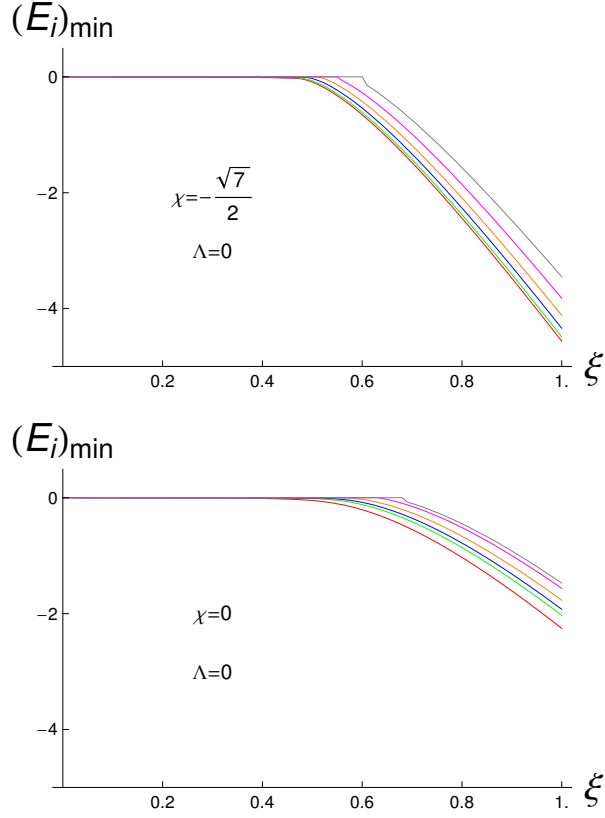}
\caption{Minimum total energies, $E_{i}(N;\beta ,\gamma ;\chi )$ as 
functions of the control parameter $\xi $, in the U(5)-SU(3) transition 
($\chi =-\frac{\sqrt{7}}{2}$) (top part) and in the U(5)-SO(6) 
transition ($\chi =0$) (bottom part), both with $\Lambda =0$.}
\label{fig15}
\end{figure}
\begin{figure}[t!]
\centering
\includegraphics[scale=0.6]{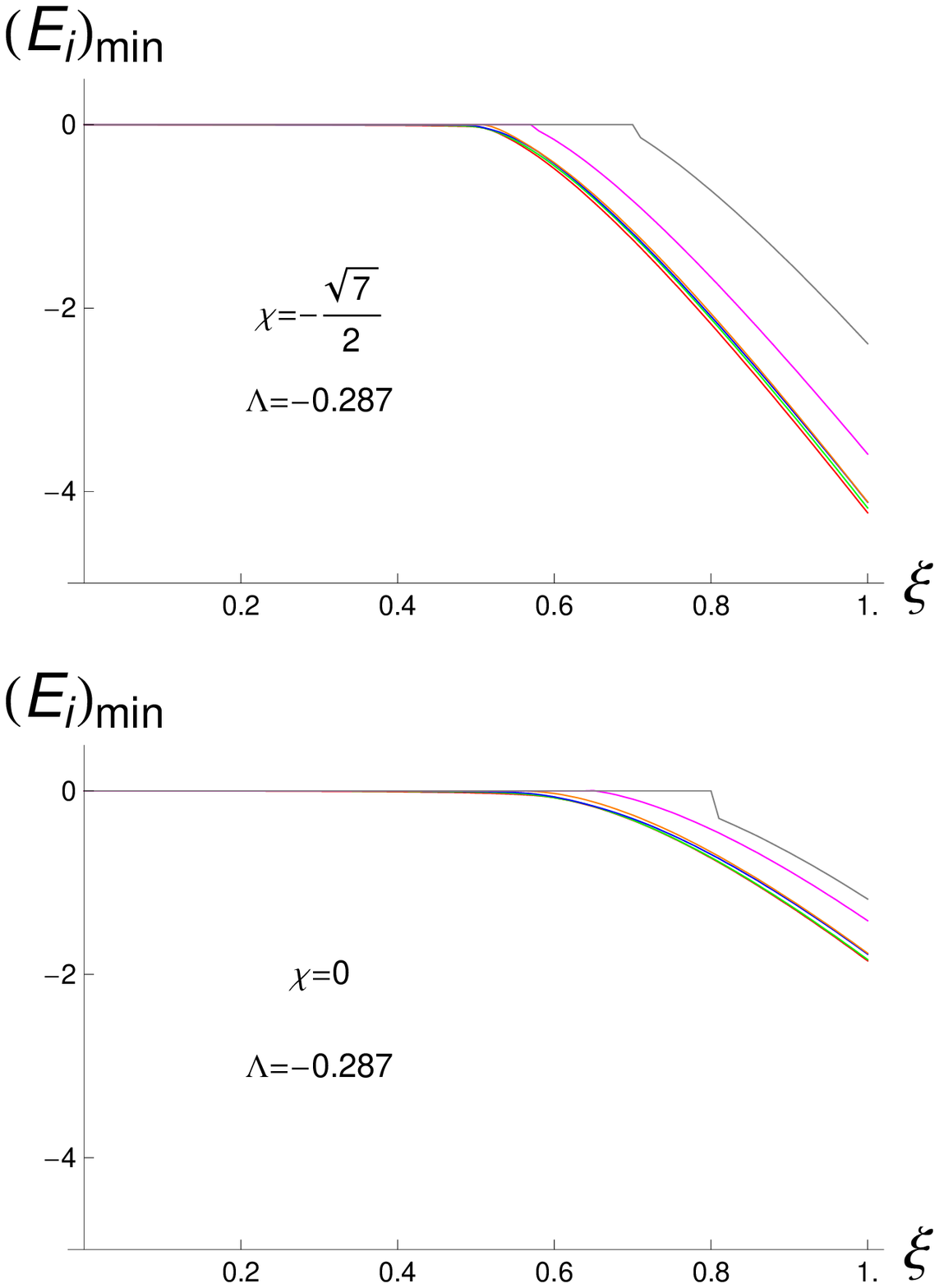}
\caption{Same as in Fig.~\ref{fig15} but for $\Lambda =-0.287$.}
\label{fig16}
\end{figure}

\subsubsection{Results for $\Lambda \neq 0$}

Results for $\chi =-\frac{\sqrt{7}}{2}$, $\chi =0$ and $\Lambda =-0.287$
are shown in Fig.~\ref{fig14}. 
The single-particle energies are more spread in this case than when 
$\Lambda=0$ and have a different ordering.

\subsection{Total energies as a function of $\protect\xi $}
\label{subSec3.6}

It is of interest to show also the results for the total energies
$E_{i}(N;\beta _{e,i},\gamma _{e,i})$ 
as a function of $\xi$. (The total energies depend also on 
$\chi$ and $\Lambda$). Throughout this section we set the scale parameter 
$\varepsilon_{0}=1$ and $N=10$. 

\subsubsection{Results for $\Lambda =0$}

The effects of the odd particle in the total ground state energies are
small, of order $1/N$, as shown in Fig.~\ref{fig15}.

\subsubsection{Results for $\Lambda \neq 0$}

The effects of the odd particle are also small here, as shown in 
Fig~\ref{fig16}, for $\Lambda =-0.287$, 
although larger than in the case of $\Lambda =0$ since the reordering of the
single particle states produces a lowering of some of them.

\section{Quantal analysis}
\label{Sec4}

The crucial aspect in the study of quantum phase transitions is obviously
the quantal analysis. This analysis must match the classical analysis of the
previous section, valid in the limit $N\rightarrow \infty $. The quantal
analysis for IBFM is done by diagonalizing numerically\ the Hamiltonian 
\begin{equation}
H=H_{B}+H_{F}+V_{BF}
\label{Eq:hIBFM2}
\end{equation}
for finite $N$ using the program ODDA~\cite{scholtenodda}. We take in this
study $N=10$. Although this is a relatively small value, previous studies in
the purely bosonic case have shown that the salient features of QPT are
already apparent~\cite{zamfir} at $N=10$.

The program ODDA uses a semi-microscopic version of IBFM in which (for a
single fermion with angular momentum~$j$)
\begin{eqnarray}
H_{F} &=& \varepsilon _{j}\,\hat{n}_{j} 
\nonumber\\
V_{BF} &=& V_{BF}^{MON}+V_{BF}^{QUAD}+V_{BF}^{EXC}
\label{hFvBF2}
\end{eqnarray}
and
\begin{eqnarray}
V_{BF}^{MON} &=&A_{s}\,\hat{n}_{d}\,\hat{n}_{j}  
\nonumber \\
V_{BF}^{QUAD} &=&
\Gamma_{s}\,\left( u_{j}^{2}-v_{j}^{2}\right) 
Q_{jj}\left[ \hat{Q}^{\chi }\cdot \hat{q}_{j} 
+ \hat{q}_{j}\cdot \hat{Q}^{\chi }\right] 
\nonumber\\
V_{BF}^{EXC} &=&-\Lambda _{s}\,8\sqrt{5}u_{j}^{2}v_{j}^{2}Q_{jj}^{2}\frac{1}{
\sqrt{2j+1}}:\left[ \left( d^{\dag }\times \tilde{a}_{j}\right) ^{(j)}\times
\left( \tilde{d}\times a_{j}^{\dag }\right) ^{(j)}\right] _{0}^{(0)}: 
\label{Eq:hIBFMcalc}
\end{eqnarray}
with
$Q_{jj}$ defined in Eq.~(\ref{Eq:Qjj}). In this semi-microscopic version, 
the Bose-Fermi interaction is given in terms of the 
BCS occupation probabilities, $u_{j}$ and $v_{j}$, with $
u_{j}^{2}+v_{j}^{2}=1$. Comparing with the model Hamiltonian~(\ref{Eq:MON}) 
of Section~\ref{Sec2}, we have the relationships 
\begin{eqnarray}
A &=&A_{s}  
\nonumber \\
\Gamma &=&\Gamma _{s}\, 2\left( u_{j}^{2}-v_{j}^{2}\right) Q_{jj}  
\nonumber \\
\Lambda &=&-\Lambda _{s}\, 8\sqrt{5}u_{j}^{2}v_{j}^{2}Q_{jj}^{2}/(2j+1) ~.
\label{Eq:IBFMcalcparam}
\end{eqnarray}

\subsection{The transition from spherical to axially deformed 
(U(5) to SU(3))}
\label{subSec4.1}

We study this transition by using a slightly modified form of the boson
Hamiltonian
\begin{eqnarray}
H_{B}^{U(5)-SU(3)} &=&\varepsilon _{0}
\left [\,\left( 1-\xi \right) \hat{n}_{d} 
-\frac{\xi }{4N} \left ( \hat{Q}^{\chi }\cdot \hat{Q}^{\chi }
+\frac{3}{8}\hat{L}\cdot \hat{L}\right ) \,\right ]  
\nonumber \\
&=&\varepsilon _{0}\left [\,
\left( 1-\xi \right) \hat{C}_{1}(U(5))-\frac{\xi }{4N}
\frac{1}{2}\hat{C}_{2}(SU(3))\,\right ] ~,
\label{Eq:hu5su3}
\end{eqnarray}
with $\chi =-\frac{\sqrt{7}}{2}$. In~(\ref{Eq:hu5su3}) 
a term $\hat{L}\cdot \hat{L}$ has been added to the Hamiltonian 
of Eq.~(\ref{Eq:hBhFvBF}), where $\hat{L}$ is the boson
angular momentum operator. 
The combination $2[\hat{Q}\cdot \hat{Q}+\frac{3}{8}\hat{L}\cdot \hat{L}]$ 
is then the quadratic Casimir operator of SU(3).
Also we denote by $\hat{C}_{1}(g)$ and $\hat{C}_{2}(g)$ Casimir operators of
the first and the second order of the algebra $g$. The reason for using this
boson Hamiltonian is that we want to isolate the intrinsic from the
rotational part of the spectrum, in order to make the comparison between
classical and quantal analysis straightforward. In the Bose-Fermi
interaction, we set $A_{s}=0$ for the same reason as in the previous section.

\subsubsection{Correlation diagrams}

\begin{figure}[h!]
\centering
\includegraphics[scale=0.55]{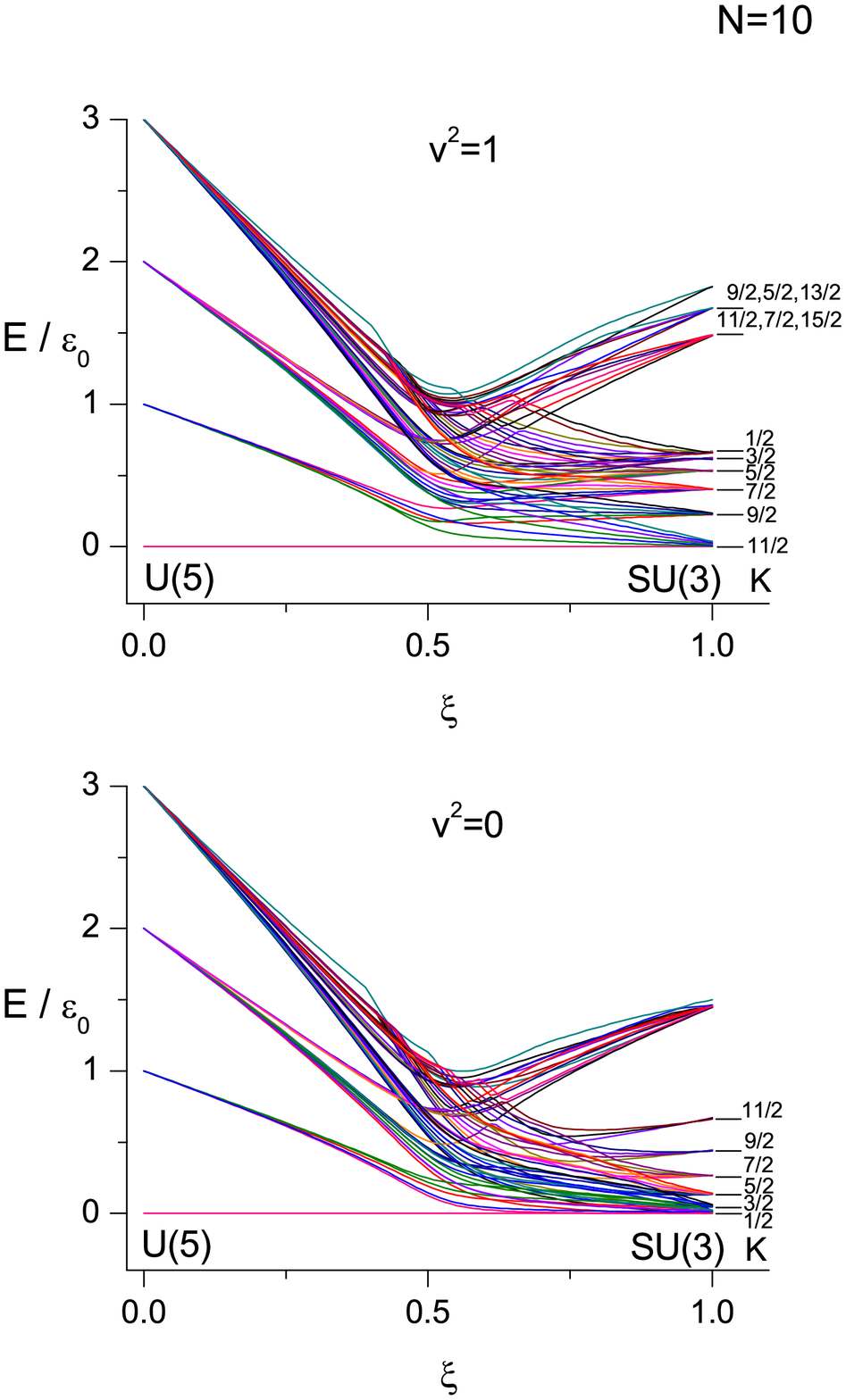}
\caption{Correlation diagram for a $j=11/2$ particle coupled to a system of 
($s,d$) bosons undergoing a U(5)-SU(3) phase transition. Top part $v^{2}=1$,
bottom part $v^{2}=0$. The interaction is purely quadrupole.}
\label{fig17}
\end{figure}

We first study the nature of the spectra in the presence of only a
quadrupole interaction, $V_{BF}^{QUAD}$, and set, 
\begin{equation}
\Gamma _{s} = \frac{\xi}{4N}\,\varepsilon _{0} ~.
\label{Eq:GammaS}
\end{equation}
In the parameterization~(\ref{Eq:hIBFMcalc}) we need to specify the values 
of the occupation probabilities, $v_{j}^{2}$. The correlation diagrams for 
$u_{j}^{2}=0,v_{j}^{2}=1$ (particle-like spectra) and 
$u_{j}^{2}=1,v_{j}^{2}=0 $ (hole-like spectra) are shown in 
Fig.~\ref{fig17}.
Correlation diagrams describe how the energy levels evolve from one phase 
to the other as a function of the control parameter $\xi $. The phases are
defined here by their symmetries, U(5) spherical, SU(3) axially deformed,
and SO(6) $\gamma $-unstable. Fig.~\ref{fig17} 
shows how rotational bands emerge from
the spherical basis. It also displays in a clear fashion particle-hole
conjugation, that is the transformation ($u\longleftrightarrow v$).

\begin{figure}[h!]
\centering
\includegraphics[scale=0.45]{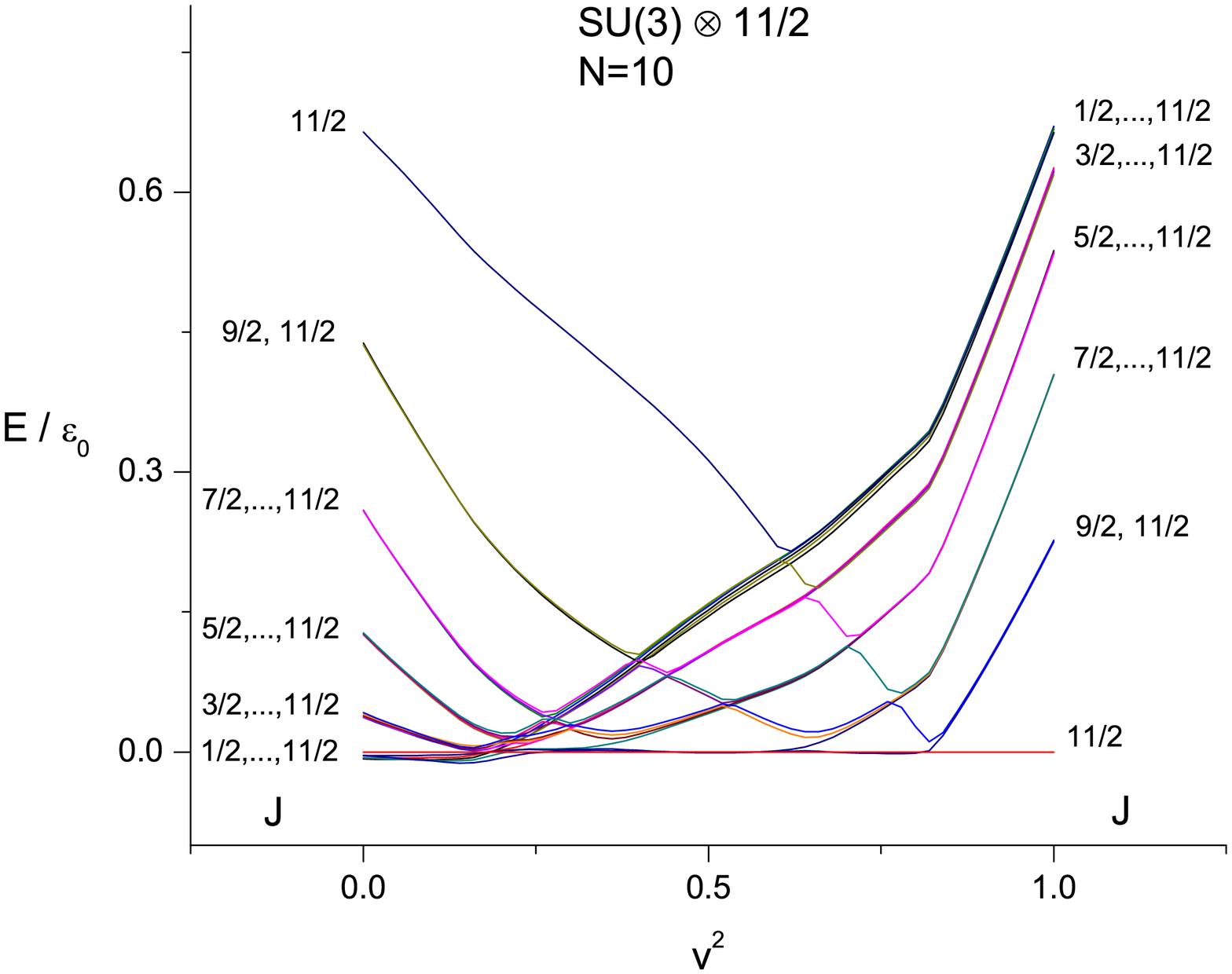}
\caption{Correlation diagram for a $j=11/2$ particle coupled to a system 
of ($s,d$) bosons with quadrupole, 
$V_{BF}^{QUAD},\chi =-\frac{\sqrt{7}}{2}$, and
exchange, $V_{BF}^{EXC}$, interaction parametrized as in 
Eq.~(\ref{Eq:hIBFMcalc}), with $\Lambda_{s}/\Gamma_{s}= 3$, 
as a function of the filling 
probability $v^{2}\equiv v_{j}^{2}$ of the single
particle state. States are labelled by the total angular momentum~$J$.}
\label{fig18}
\end{figure}

With the values of the BCS coefficients used in Fig.~\ref{fig17}, 
the exchange interaction vanishes. 
To study the effect of the exchange interaction, we
construct the correlation diagram as a function of $v_{j}^{2}$, 
Fig.~\ref{fig18}. 
This is the same as varying $\Lambda $, since $u_{j}^{2}=1-v_{j}^{2}$ and
thus 
\begin{equation}
\Lambda =
-\Lambda _{s}\, 8\sqrt{5}(1-v_{j}^{2})v_{j}^{2}Q_{jj}^{2}/(2j+1) ~.
\label{Eq:lamlamS}
\end{equation}
As $v_{j}^{2}$ changes from $0$ to $1$, we span the entire set of spectra
from hole-like to particle-like. Particularly interesting in this figure are
the spectra at half-filling, $v_{j}^{2}=0.5$. At this point, the quadrupole
interaction vanishes, and the spectrum is given by the exchange interaction.
It should be noted, however, that the diagram shown in Fig.~\ref{fig18} is 
an unconventional correlation diagram, since by varying $v_{j}^{2}$ we vary
both the exchange and the quadrupole interaction,~Eq.~(\ref{Eq:hIBFMcalc}).

\subsubsection{Classical-quantal correspondence}

In addition to providing a description of the evolution of the quantal
levels from one symmetry (phase) to another, correlation diagrams are also
useful to test the classical-quantal correspondence. In Section~\ref{Sec3}, 
the single-particle energies for a fermion with angular momentum~$j$ in the
presence of the boson condensate of Eq.~(\ref{Eq:bc}) were calculated 
classically. For $\gamma =0^{\circ }$ they were given by 
Eqs.~(\ref{Eq:lamK}) and (\ref{Eq:Pj}), as a function of $
\beta $ and for any $\chi $. When the bosons have SU(3) symmetry, $\chi =-
\frac{\sqrt{7}}{2}$, and in the limit $N\rightarrow \infty $, the
equilibrium deformation is $\beta _{e}=\sqrt{2}$. Inserting these values in
Eq.~(\ref{Eq:lamK}) we obtain the classical result
\begin{eqnarray}
\lambda _{K}^{SU(3)} &=&
-N\Gamma \sqrt{2}\sqrt{5}P_{j}\left[ 3K^{2}-j\left(j+1\right) \right]   
\nonumber \\
&& 
-N\Lambda \frac{2}{3}\left( 2j+1\right) 
P_{j}^{2}\left[ 3K^{2}-j(j+1)\right] ^{2} ~.
\label{Eq:lamKsu3}
\end{eqnarray}
An analytic quantal calculation of the single-particle energies for bosons
with SU(3) symmetry in the limit of large $N$ can also be done. The bosonic
ground state wave function, with good SU(3) quantum numbers, 
can be written as 
$\left\vert \left[ N\right] ,\left( 2N,0\right) ,
K_{c}=0,L,M\right\rangle$~\cite{iac111}. 
States for a fermion with angular momentum~$j$ coupled to the bosonic 
ground state can be written as 
$\left\vert \left[ N\right] ,(2N,0),K_{c}=0;j,K_{j};K=K_{j},J,M
\right\rangle$. 
A long but straightforward calculation of the matrix elements of the
interaction $V_{BF} = V_{BF}^{QUAD}+V_{BF}^{EXC}$, 
with $V_{BF}^{QUAD}$ and $V_{BF}^{EXC}$ as in Eq.~(\ref{Eq:hIBFMcalc}), 
gives~\cite{scholtenthesis}
\begin{equation}
\lim_{N\rightarrow\infty}
\left\langle \left[ N\right] ,\left( 2N,0\right) ,0;j,K_{j};K,J,M
\mid V_{BF}\mid 
\left[ N\right] ,\left( 2N,0\right) ,0;j,K_{j};K,J,M\right\rangle 
= \lambda _{K}^{SU(3)} ~.
\label{Eq:vBFsu3}
\end{equation}
Here the parameters $\Gamma$ and $\Lambda$ in the expression 
for $\lambda _{K}^{SU(3)}$, Eq.~(\ref{Eq:lamKsu3}), are related to 
the parameters of $V_{BF}$, Eq.~(\ref{Eq:hIBFMcalc}), 
by means of Eqs.~(\ref{Eq:IBFMcalcparam}) and (\ref{Eq:GammaS}). 
Specifically, 
$\Gamma = 2\Gamma_{s}(1-2v_{j}^2)Q_{jj}$, 
$\Gamma _{s} = \frac{\xi}{4N}\,\varepsilon _{0}$ 
and $\Lambda$ is given in Eq.~(\ref{Eq:lamlamS}). 
It should also be noted that in the limit $N\rightarrow \infty $, 
the expectation value of $V_{BF}$ is independent of the the total 
angular momentum~$J$.

Both the explicit formula Eq.~(\ref{Eq:vBFsu3}) and the independence 
on~$J$ can be checked numerically. 
In Fig.~\ref{fig17} top, obtained with 
$\Gamma =-\frac{\xi }{4N}\left(2\varepsilon _{0}\right)Q_{jj} $ 
and $\Lambda =0$, one can see that at $\xi =1$,
SU(3) symmetry, all rotational levels built on the intrinsic state $K$,
converge to a single value, as seen on the right-hand side of the figure,
and that the energies of the intrinsic states can be described to a good
approximation by Eq.~(\ref{Eq:lamKsu3}). 
Since the figure represents the results for $N=10$, 
one can see that small deviations occur, especially for the band with 
$K=1/2$, but these deviations are relatively small and the large $N$ limit
appears to be already reached. The same conclusion applies to 
Fig.~\ref{fig17} bottom obtained with 
$\Gamma =+\frac{\xi }{4N}\left( 2\varepsilon _{0}\right)Q_{jj}$,
and $\Lambda =0$. Flipping the sign of $\Gamma$ (particle-hole conjugation)
reverses the ordering of the $K$ states but the energies are still given by
Eq.~(\ref{Eq:lamKsu3}). 
(The bottom part of Fig.~\ref{fig17} is also consistent with 
the top part of Fig.~\ref{fig3}, once the relationship 
$\Gamma =2\Gamma _{s}Q_{jj}$ is taken into account).

The correspondence between classical and quantal calculation is exact at 
$\xi =1$, SU(3) symmetry, and $N\rightarrow \infty $. As one moves away from 
$\xi =1$, or for finite $N$, it becomes approximate. Nonetheless, rotational
bands can still be identified for $\xi _{c}\leq \xi \leq 1$. Below the
critical value, the characterization of states by a $K$ quantum number is no
longer possible. For this region, $0\leq \xi \leq \xi _{c}$, a
straightforward comparison between classical and quantal calculation is no
longer possible and one must resort to a numerical calculation both
classical, Section~\ref{Sec3}, and quantum-mechanical, Section~\ref{Sec4}.

\subsubsection{Ground state energy}
\label{subsubSec4.1.3}

\begin{figure}[t!]
\centering
\includegraphics[scale=0.5]{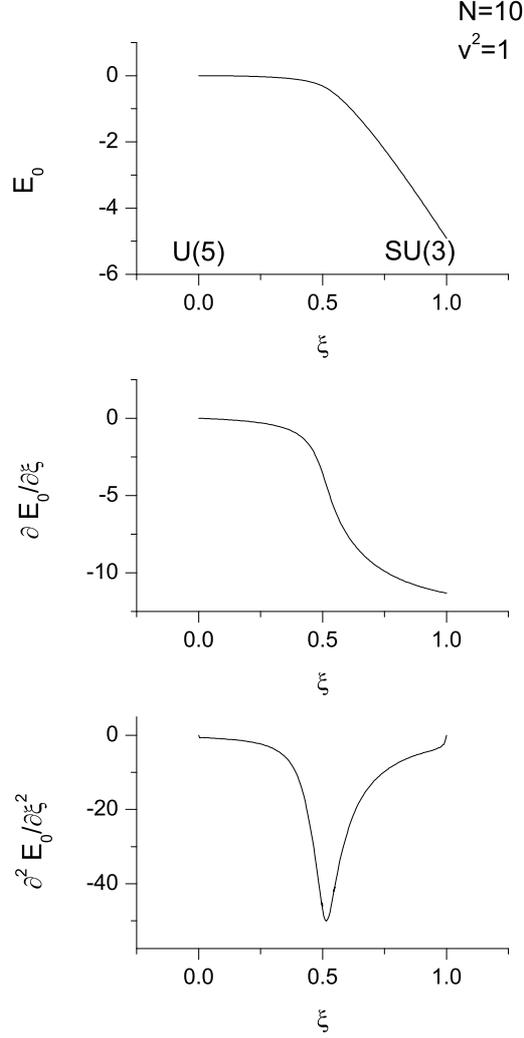}
\caption{The ground state energy, $E_{0}$, top, its first derivative, 
$\frac{\partial E_{0}}{\partial \xi }$, center, and its second 
derivative, $\frac{\partial ^{2}E_{0}}{\partial \xi ^{2}}$, bottom, 
for a $j=11/2$ particle coupled to a system of ($s,d$) bosons 
undergoing a U(5)-SU(3) transition.}
\label{fig19}
\end{figure}

The ground state energy is a key indicator of phase transitions. The ground
state energy, $E_{0}$, its first, $\frac{\partial E_{0}}{\partial \xi }$,
and second, $\frac{\partial ^{2}E_{0}}{\partial \xi ^{2}}$, derivatives with
respect to $\xi $ are shown in Fig.~\ref{fig19} (particle-like, 
$v_{j}^{2}=1$). (The energy $E_{0}$ in this figure is in units of the 
scale factor $\varepsilon_{0}$, taken to be $\varepsilon _{0}=1$ 
and $N=10$). 
For hole-like spectra, $v_{j}^{2}=0$, we have a similar behavior, as well as
in the case in which $\Lambda \neq 0$. The quantal result for $E_{0}$ should
be compared with the classical result, ($E_{0}$)$_{\min }$, shown in the top
part of Fig.~\ref{fig15}, where ($E_{0}$)$_{\min }$ is the energy of the 
lowest state at the minimum. Within $1/N$ corrections, the two results 
agree (classical-quantal correspondence). Fig.~\ref{fig19} is similar to 
Fig.~(7.3) of~\cite{williams} for the purely bosonic case. 
Although the value of $N$ used here is too small to distinguish between 
first and second order transition, with discontinuities in 
$\frac{\partial E_{0}}{\partial \xi }$ and 
$\frac{\partial ^{2}E_{0}}{\partial \xi ^{2}}$ respectively, 
nonetheless precursors of the QPT are clearly seen at 
$\xi =\xi _{c}\cong 0.5$.

\subsubsection{Quantal order parameters}

\begin{figure}[t!]
\centering
\includegraphics[scale=0.5]{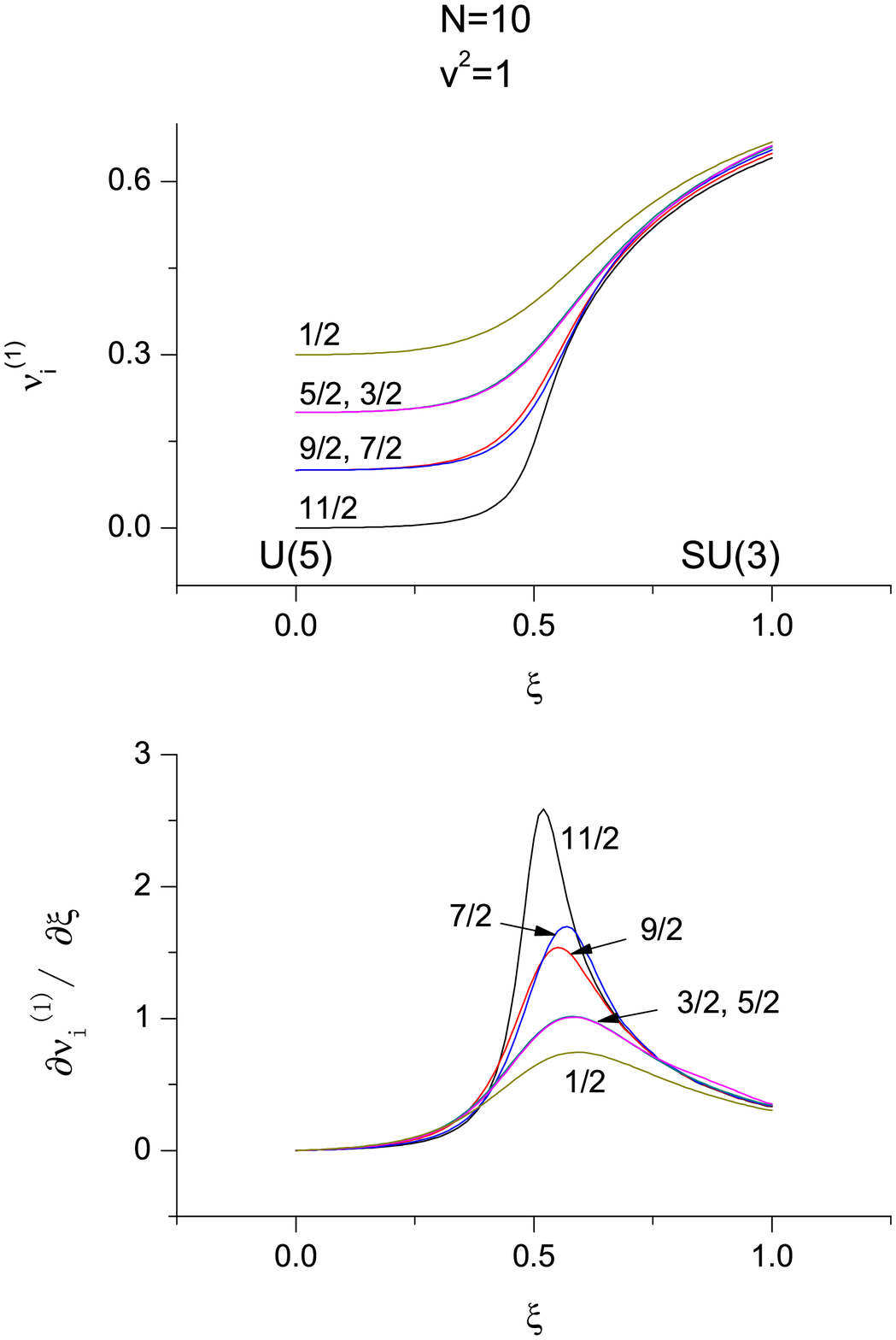}
\caption{The quantal order parameters, $\nu_{i}^{(1)}$, 
Eq.~(\ref{Eq:orderparam1}), as a
function of the control parameter, $\xi $ (top part). The expectation values
of the number of $d$ bosons, $\left\langle \hat{n}_{d}\right\rangle $, for
the lowest states of a given total angular momentum, $J=11/2,...,1/2$ for a 
$j=11/2$ particle coupled to a system of ($s,d$) bosons undergoing a
U(5)-SU(3) transition. Particle-like spectra, $v^{2}=1$, and no exchange
interaction. The derivative $\frac{\partial\nu _{i}^{(1)}}{\partial \xi }$
as a function of $\xi $ (bottom part).}
\label{fig20}
\end{figure}

The equilibrium deformations $\beta _{e,i},\gamma _{e,i}$ are the classical
order parameters. As quantal order parameters we consider here only the
expectation value of $\hat{n}_{d}$ in the states 
$i=1,...,6$, (first quantal order parameter),~$\nu _{i}^{(1)}$
\begin{equation}
\nu _{i}^{(1)} = \frac{\left\langle \psi _{i}
\left\vert \hat{n}_{d}\right\vert\psi _{i}\right\rangle }{N} ~.
\label{Eq:orderparam1}
\end{equation}
This is shown in Fig.~\ref{fig20} top part.
Since the order parameters $\nu _{i}^{(1)}$ are related to the square of the
classical order parameters $\beta _{e,i}$, this figure is related to 
Fig.~\ref{fig9} to which it corresponds in the limit $N\rightarrow \infty $. 
The derivative of $\nu _{1}^{(1)}$ in the ground state, 
$\frac{\partial \nu _{i}^{(1)}}{\partial \xi }$ is also shown in 
Fig.~\ref{fig20} bottom part. (This quantity
diverges when $N\rightarrow \infty $). From this figure one sees clearly
that the transition is made sharper by the presence of the fermion for some
states, $11/2,9/2,7/2$, while is made smoother for others, $5/2,3/2,1/2$, a
result already seen in the classical analysis.

\subsection{The transition from spherical to $\protect\gamma -$unstable
(U(5)-SO(6))}
\label{subSec4.2}

We study this transition by using the standard form of the transitional
boson Hamiltonian
\begin{equation}
H_{B}^{U(5)-SO(6)}=\varepsilon _{0}\left [
\left( 1-\xi \right)\hat{n}_{d}
-\frac{\xi }{4N}\,\hat{Q}^{\chi=0}\cdot \hat{Q}^{\chi=0}\,\right ]
\label{Eq:hu5so6}
\end{equation}
with $\chi =0$. We set $A_{s}=0$ as in the previous section and 
$\Gamma_{s}=\frac{\xi }{4N}\varepsilon _{0}$ as in Eq.~(\ref{Eq:GammaS}).

\subsubsection{Correlation diagram}

\begin{figure}[t!]
\centering
\includegraphics[scale=0.47]{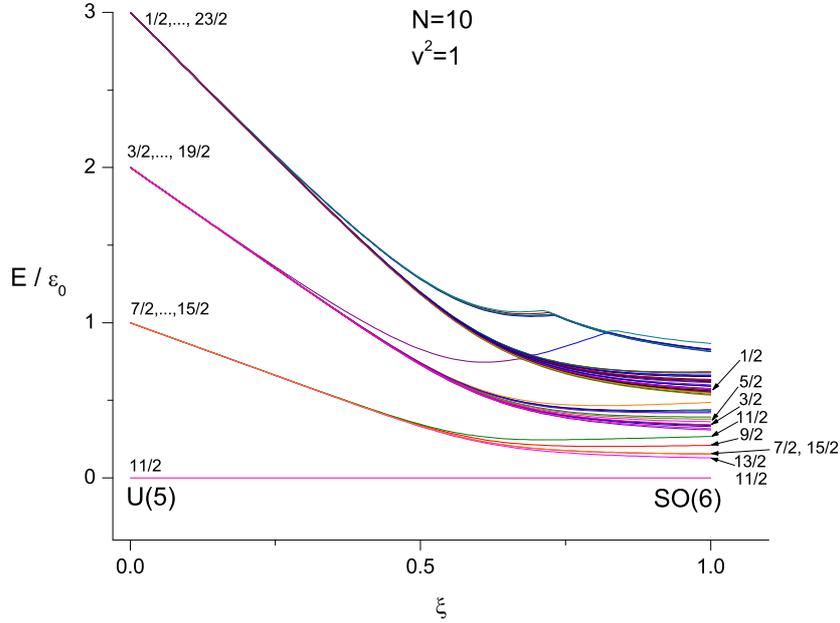}
\caption{Correlation diagram for a $j=11/2$ particle coupled to a system of 
($s,d$) bosons undergoing an U(5)-SO(6) transition. The interaction is 
purely quadrupole. There is no difference in this case between 
particle-like ($v^{2}=1$) and hole-like ($v^{2}=0$) spectra.}
\label{fig21}
\end{figure}

The correlation diagram for this phase transition is shown in 
Fig.~\ref{fig21}.
\begin{figure}[h!]
\centering
\includegraphics[scale=0.45]{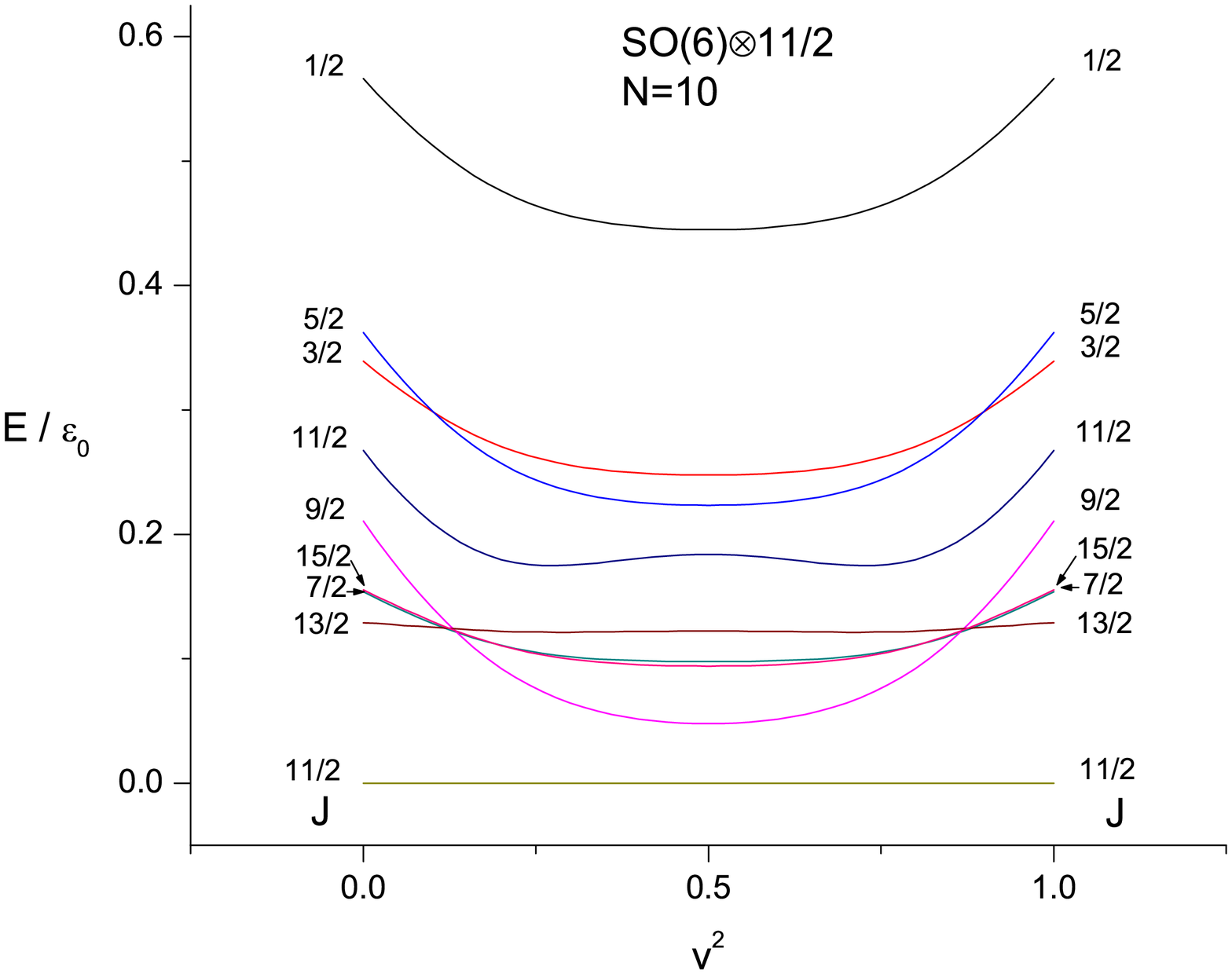}
\caption{Same as Fig.~\ref{fig18} but with $\chi =0$.}
\label{fig22}
\end{figure}
In this case, there is no clear separation between intrinsic and rotational
motion, since $\hat{Q}^{\chi=0 }\cdot \hat{Q}^{\chi=0 }$ is not 
a Casimir operator of SO(6), and thus the states at $\xi =1$ do not coalesce 
into a single point as in Fig.~\ref{fig17}. 
(We have also done a study in which $\hat{C}_{2}(SO(6))$ is used instead 
of $\hat{Q}^{\chi=0 }\cdot \hat{Q}^{\chi=0 }$. These two operators are related 
by $\hat{C}_{2}(SO(6))=\hat{Q}^{\chi =0}\cdot \hat{Q}^{\chi =0}
+\hat{C}_{2}(SO(5))$. However, using the Casimir operator, makes
the figure at the end point $\xi =1$ very crowded, since all states arising
from a given $SO(6)$ representation collapse to zero energy, and, for this
reason, we prefer to plot in Fig.~\ref{fig21} the results with the 
Hamiltonian~(\ref{Eq:hu5so6}).

The effect of the exchange interaction is studied by constructing the
correlation diagram as a function of $v_{j}^{2}$, Fig.~\ref{fig22}. 
A remarkable property of this diagram is the symmetry under particle-hole
conjugation. The diagram is symmetric around half-filling. This is unlike
the case of the U(5)-SU(3) transition.

\subsubsection{Ground state energy}

\begin{figure}[h!]
\centering
\includegraphics[scale=0.5]{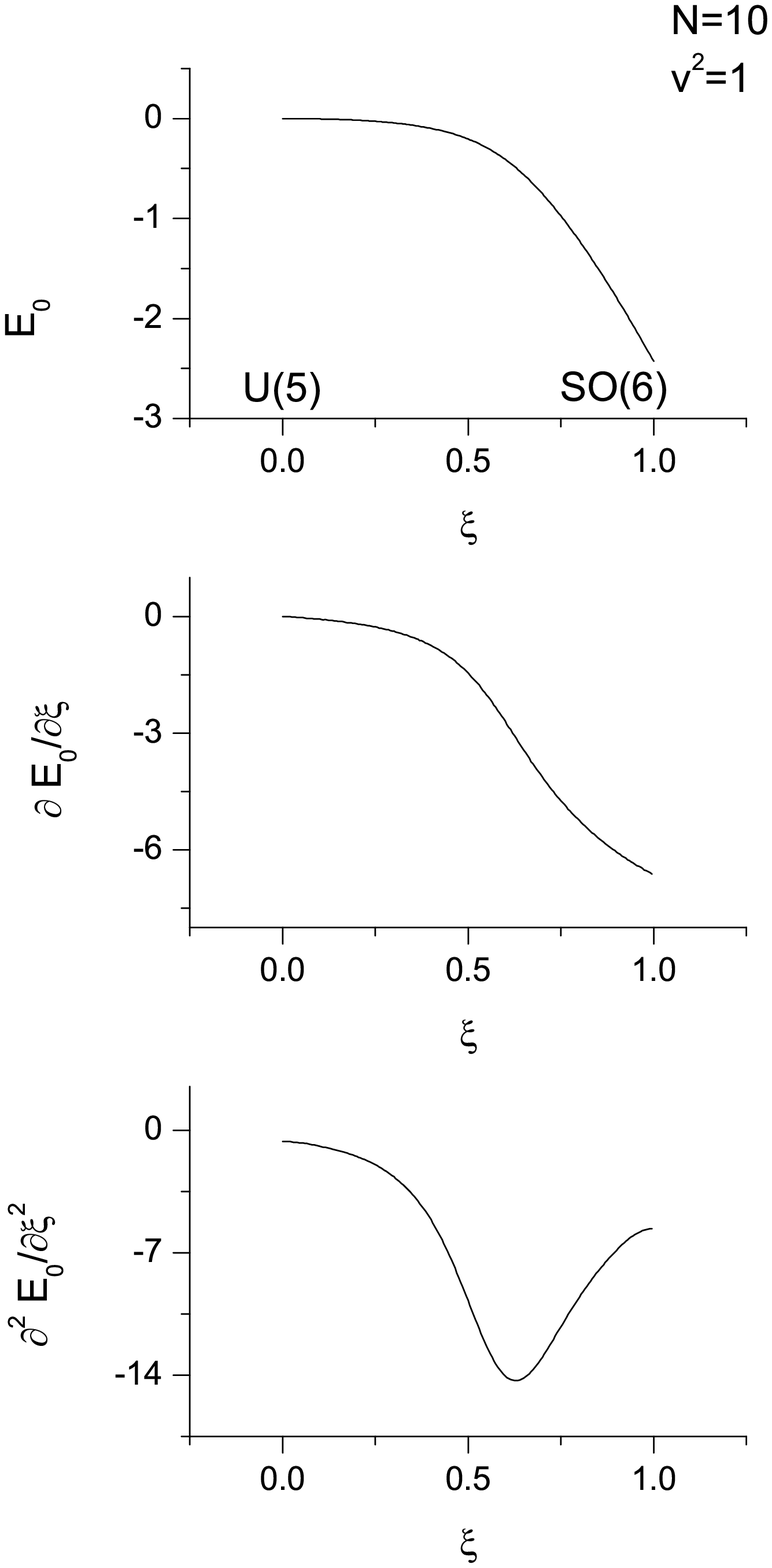}
\caption{Same as Fig.~\ref{fig19} but for the U(5)-SO(6) transition.}
\label{fig23}
\end{figure}

The ground state energy,\ $E_{0}$, its first and second derivative are 
shown in Fig.~\ref{fig23}, U(5)-SO(6) transition.
This figure is similar to Fig.~\ref{fig19}, U(5)-SU(3) transition, 
and as mentioned in Section~\ref{subsubSec4.1.3}, 
it is not possible to distinguish whether the transition is 
first or second order. In order to do so, one must go to much larger values
of $N$, as done in the purely bosonic case~\cite{werner}. However, the fact
that the quantity $\frac{\partial ^{2}E_{0}}{\partial \xi ^{2}}$ is smoother
in Fig.~\ref{fig23} than in Fig.~\ref{fig19} supports the conclusion 
that the ground state
transition is first order for U(5)-SU(3) and second order for U(5)-SO(6),
since the second derivative of $E_{0}$ diverges (or not) for first (or
second) order transitions in the limit $N\rightarrow \infty $.

\section{Experimental evidence}
\label{Sec5}

Even-even nuclei in the mass region A$\sim $150 are known to experience a
first order quantum phase transition (U(5)-SU(3)) at neutron number 90. The
evidence for this statement comes from the analysis of (i) the excitation
spectrum which displays a gap, $\Delta = E(0^{+}_2)-E(0^{+}_1)$; 
(ii) the two neutron separation
energies, $S_{2n}(N)=-[E_{0}(N+1)-E_{0}(N)]$, proportional to the derivative
of the ground state energy, $E_{0}$, with respect to the control parameter, 
$\xi $, i.e. $\frac{\partial E_{0}}{\partial \xi }$, and (iii) the 
$B(E2;0\rightarrow 2)$ values, proportional to the square of 
the order parameter 
$\left\langle \hat{n}_{d}\right\rangle^{2}$~\cite{exp2,cejnar10,reviews1}. 
In odd-even nuclei, the first two quantities can be easily measured, while
the last quantity, $B(E2;J_{g.s.}\rightarrow J^{\prime })$ is more difficult
to measure due to the fragmentation of the $B(E2)$ strength from the ground
state to several states $J^{\prime }$. In this paper, we therefore analyze
the first two quantities and show that both display the features expected
for a QPT, as described in Section~\ref{Sec3} and Section~\ref{Sec4}.

\subsection{Excitation spectrum}
\label{subSec5.1}

Odd-proton nuclei in the region of A$\sim $150 offer an unique opportunity
to study QPT in Bose-Fermi systems, because of the occurrence of the unique
parity state $h_{11/2}$ near the Fermi surface. We therefore analyze the
negative parity states of the odd-proton nuclei, $_{61}$Pm$_{86-92}$, 
$_{63}$Eu$_{86-92}$, and $_{65}$Tb$_{86-92}$. A realistic calculation for 
these nuclei can be done within the framework of IBFM, with 
an Hamiltonian $H=H_{B}+H_{F}+V_{BF}$. 
The IBM Hamiltonian $H_{B}$ is written as~\cite{iac111}
\begin{eqnarray}
H_{B} &=&\varepsilon \hat{n}_{d}+\sum_{L=0,2,4}c_{L}\frac{1}{2}\sqrt{2L+1}
\left[ (d^{\dag }\times d^{\dag })^{(L)}\times \left( \tilde{d}
\times \tilde{d}\right) ^{(L)}\right] _{0}^{(0)}  
\nonumber \\
&&+\frac{1}{\sqrt{2}}v_{2}\left[ \left( d^{\dag }\times d^{\dag }\right)
^{(2)}\times \left( \tilde{d}\times s\right) ^{(2)}
+\left( s^{\dag }\times d^{\dag }\right) ^{(2)}
\times \left( \tilde{d}\times \tilde{d}\right) ^{(2)}\right] _{0}^{(0)}  
\nonumber \\
&&+\frac{1}{2}v_{0}\left[ \left( d^{\dag }\times d^{\dag }\right)
^{(0)}\times \left( s\times s\right) ^{(0)}
+\left( s^{\dag }\times s^{\dag}\right) ^{(0)}\times 
\left( \tilde{d}\times \tilde{d}\right) ^{(0)}\right]_{0}^{(0)} ~.
\label{Eq:hIBM}
\end{eqnarray}
In order to perform a calculation of the odd proton nuclei indicated above,
we need the Hamiltonian parameters of the even-even nuclei $_{60}$Nd$_{86-92}
$, $_{62}$Sm$_{86-92}$, $_{64}$Gd$_{86-92}$. We take these parameters from
previous studies. In these studies, first a calculation in the
Proton-Neutron Interacting Boson Model (IBM-2) is done with parameters given
in Table~\ref{Table1}.
\begin{table}
\begin{center}
\begin{tabular}{|c|cccccccc|}
\hline
& & & & & & & &\\[-3mm] 
& $N_{\pi }$ & $N_{\nu }$ & $\varepsilon $ & $\kappa $ & $\chi _{\pi }$ 
& $\chi _{\nu }$ & $c_{0}^{\pi }$ & $c_{2}^{\pi }$ \\
& & & & & & & &\\[-3mm] 
\hline
& & & & & & & &\\[-3mm] 
$^{146}$Nd & 5 & 2 & 0.90 & -0.150 & -1.2 & 0.0 & 0.4 & 0.2 \\[2pt] 
$^{148}$Nd & 5 & 3 & 0.73 & -0.100 & -1.2 & -0.8 & 0.4 & 0.2 \\[2pt]
$^{150}$Nd & 5 & 4 & 0.48 & -0.070 & -1.2 & -1.0 & 0.4 & 0.2 \\[2pt] 
$^{152}$Nd & 5 & 5 & 0.37 & -0.089 & -1.2 & -1.1 & 0.4 & 0.2 \\[2pt] 
\hline
& & & & & & & &\\[-3mm] 
$^{148}$Sm & 6 & 2 & 0.90 & -0.120 & -1.3 & 0.0 & 0.0 & 0.05 \\[2pt] 
$^{150}$Sm & 6 & 3 & 0.70 & -0.076 & -1.3 & -0.8 & 0.0 & 0.05 \\[2pt]
$^{152}$Sm & 6 & 4 & 0.52 & -0.071 & -1.3 & -1.0 & 0.0 & 0.05 \\[2pt] 
$^{154}$Sm & 6 & 5 & 0.44 & -0.079 & -1.3 & -1.1 & 0.0 & 0.05 \\[2pt] 
\hline
& & & & & & & &\\[-3mm] 
$^{150}$Gd & 7 & 2 & 0.95 & -0.090 & -1.0 & 0.0 & -0.2 & -0.1 \\[2pt] 
$^{152}$Gd & 7 & 3 & 0.70 & -0.070 & -1.0 & -0.8 & -0.2 & -0.1 \\[2pt] 
$^{154}$Gd & 7 & 4 & 0.55 & -0.072 & -1.0 & -1.0 & -0.2 & -0.1 \\[2pt] 
$^{156}$Gd & 7 & 5 & 0.46 & -0.073 & -1.0 & -1.1 & -0.2 & -0.1 \\[4pt] 
\hline
\end{tabular}
\caption{Parameters of IBM-2 used in the calculation of the even-even
nuclei $^{146-150}$Nd, $^{148-152}$Sm, $^{150-156}$Gd.}\label{Table1}
\end{center}
\end{table}
\begin{table}
\begin{center}
\begin{tabular}{|c|cccc|}
\hline
& & & & \\ [-3mm]
& $A_{s}$ & $2\Gamma _{s}$ & $\Lambda _{s}$ & $v_{j}^{2}$ \\ [1mm]
& & & & \\ [-3mm]
\hline
& & & & \\ [-3mm]
$^{147}$Pm & -0.1 & 0.94 & 2.52 & 0.18 \\[2pt] 
$^{149}$Pm & -0.1 & 1.26 & 3.39 & 0.25 \\[2pt] 
$^{151}$Pm & -0.1 & 1.52 & 4.11 & 0.33 \\[2pt] 
$^{153}$Pm & -0.1 & 1.67 & 4.51 & 0.33 \\[2pt] 
\hline
& & & & \\ [-3mm]
$^{149}$Eu & -0.1 & 0.83 & 1.91 & 0.30 \\[2pt] 
$^{151}$Eu & -0.1 & 1.15 & 2.63 & 0.32 \\[2pt] 
$^{153}$Eu & -0.1 & 1.36 & 3.45 & 0.34 \\[2pt] 
$^{155}$Eu & -0.1 & 1.51 & 3.53 & 0.35 \\[2pt] 
\hline
& & & & \\ [-3mm]
$^{151}$Tb & -0.1 & 0.73 & 1.25 & 0.36 \\[2pt] 
$^{153}$Tb & -0.1 & 1.05 & 1.97 & 0.38 \\[2pt] 
$^{155}$Tb & -0.1 & 1.26 & 2.79 & 0.40 \\[2pt] 
$^{157}$Tb & -0.1 & 1.41 & 2.87 & 0.41 \\[4pt]
\hline
\end{tabular}
\caption{Strengths of the Bose-Fermi couplings in the odd-even nuclei $
^{147-153}$Pm, $^{149-155}$Eu, $^{151-157}$Tb.}
\label{Table2}
\end{center}
\end{table}
The input parameters in IBM, $\varepsilon ,c_{L}(L=0,2,4),v_{2},v_{0}$,
Eq.~(\ref{Eq:hIBM}) are then calculated from these by a projection 
technique~\cite{scholtenthesis}.

\begin{figure}[t]
\centering
\includegraphics[scale=0.43]{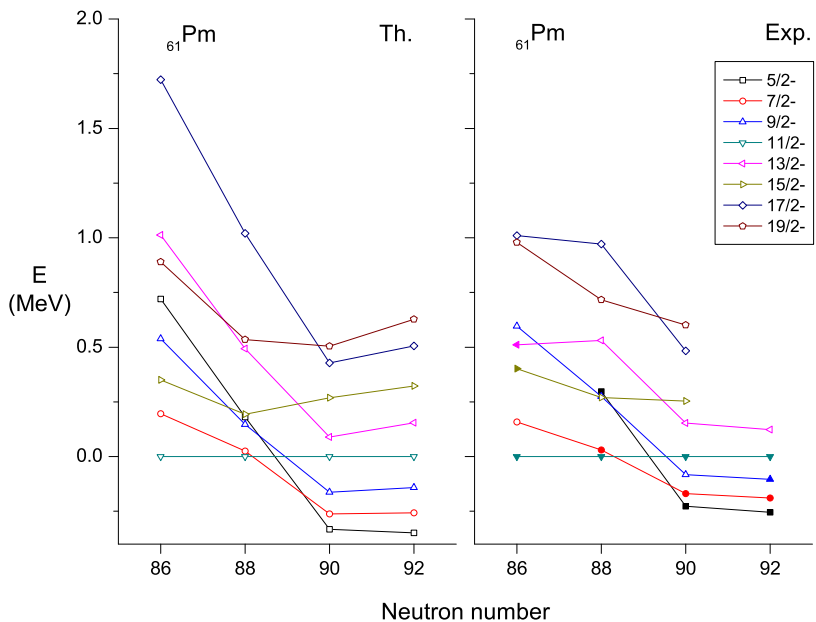}
\caption{Comparison between calculated and experimental spectra of negative
parity states in $_{61}$Pm. The lowest $11/2^{-}$ state is taken as zero of
the energy. The parameters of the calculation are given in 
Tables~\ref{Table1} and~\ref{Table2}. 
In the experimental spectra, taken from~\cite{data1}, 
uncertain assignments are indicated by open symbols.}
\label{fig24}
\end{figure}

\begin{figure}[t]
\centering
\includegraphics[scale=0.43]{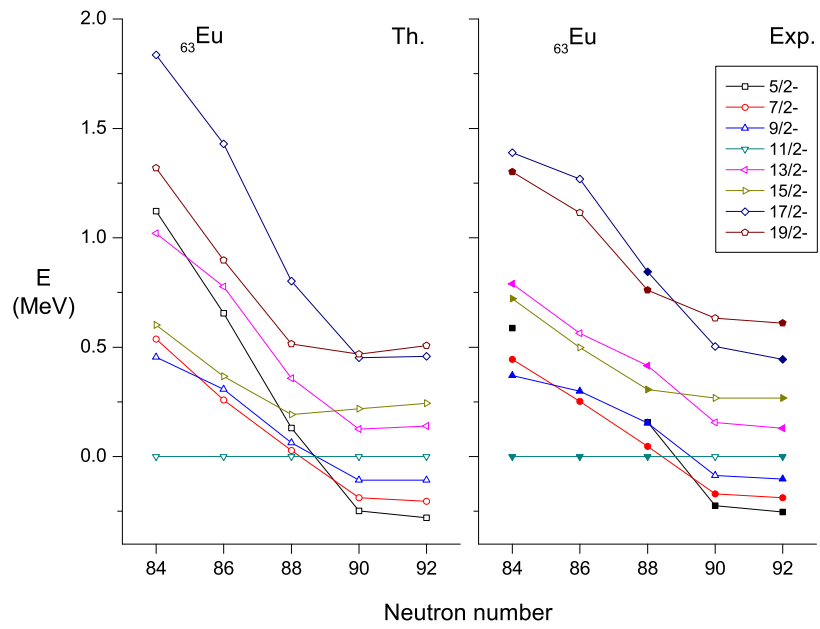}
\caption{Same as Fig.~\ref{fig24} but for $_{63}$Eu.}
\label{fig25}
\end{figure}

\begin{figure}[h!]
\centering
\includegraphics[scale=0.43]{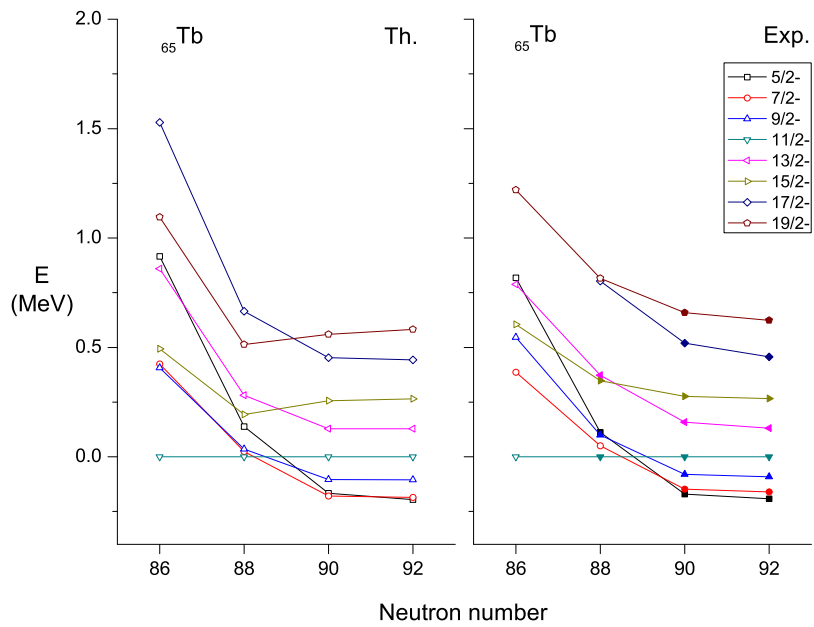}
\caption{Same as Fig.~\ref{fig25} but for $_{65}$Tb.}
\label{fig26}
\end{figure}

In the case of the unique parity configuration $h_{11/2}$ there is only one
single particle energy, $\varepsilon _{j}$, which we take as $\varepsilon
_{j}=0$. The Bose-Fermi interaction~(\ref{Eq:hIBFMcalc}) is specified 
by the parameters 
$A_{s},\Gamma _{s}$ and $\Lambda _{s}$ and $v_{j}^2$, three of which are 
independent. Using the semi-microscopic theory, one can extract the 
occupation probabilities $v_{j}^{2}$. 
The parameters used in the present analysis are given in 
Table~\ref{Table2}. For $_{63}$Eu and $_{61}$Pm isotopes, they 
are the same parameters used by Scholten and Blasi~\cite{scholten-blasi} 
and by Scholten and Ozzello~\cite{scholten-ozzello}, respectively, 
in previous studies of these isotopes. 
For the $_{65}$Tb isotopes we scale the latter parameters. 
Our calculated spectra are shown in 
Figs.~\ref{fig24},~\ref{fig25},~\ref{fig26}, where they are
compared with the available experimental data~\cite{data1}.
One can see very clearly the phase transition occurring between neutron
numbers 88 and 90 both in the theoretical (left) and in the experimental
(right) spectra. 

\subsection{Two-neutron separation energies}
\label{subSec5.2}

\begin{figure}[t]
\centering
\includegraphics[scale=0.43]{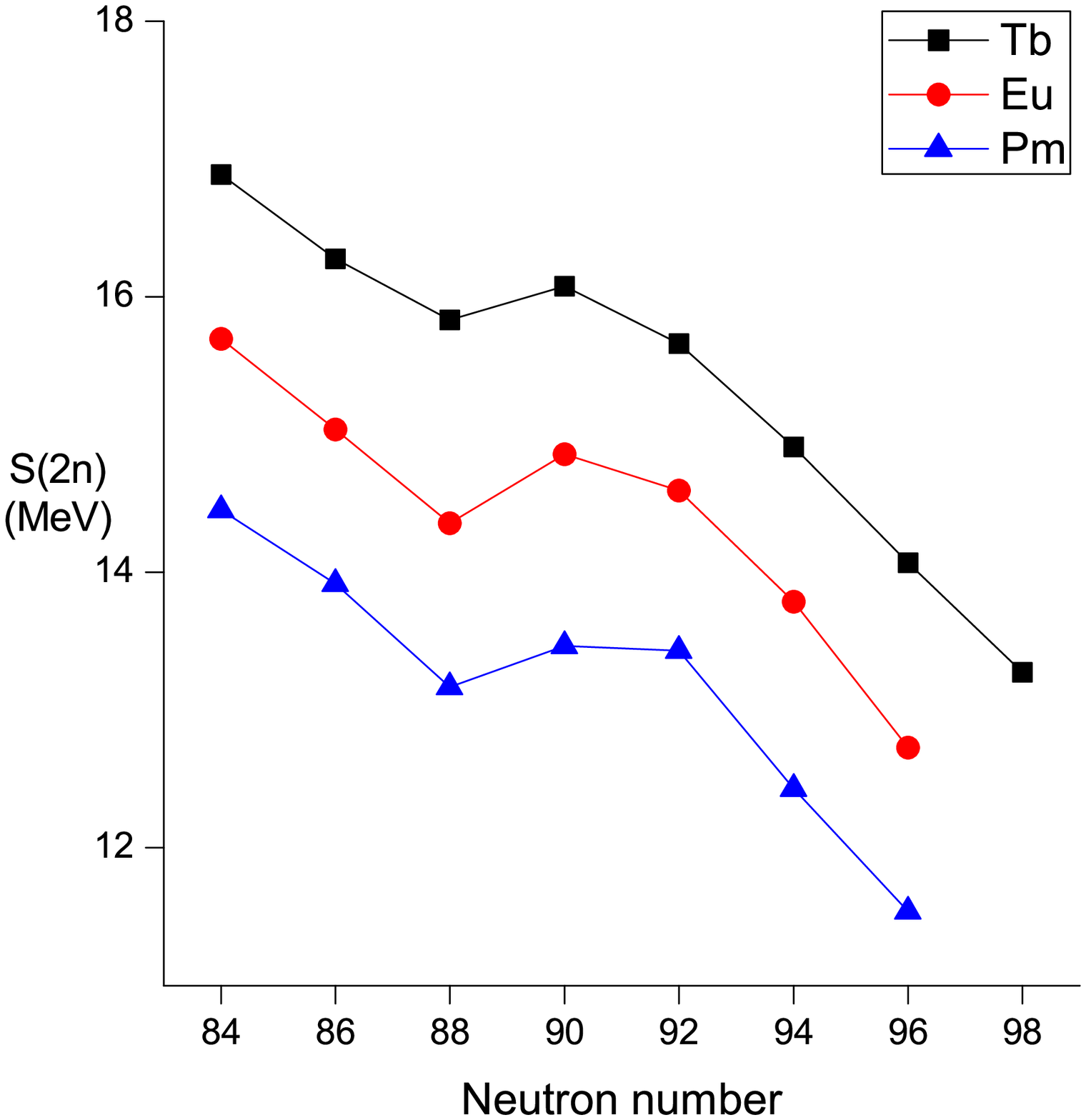}
\caption{Experimental two-neutron separation energies, $S(2n)$, for the
even-neutron isotopes of $_{61}$Pm, $_{63}$Eu and $_{65}$Tb, 
taken from~\cite{data2}.}
\label{fig27}
\end{figure}

A complete analysis of two-neutron separation energies requires a
calculation of both positive and negative parity states in $_{61}$Pm, $_{63}$
Eu and $_{65}$Tb. For the Eu isotopes this calculation was done by 
Scholten~\cite{scholten-blasi}. 
The experimental two-neutron separation energies in 
$_{61}$Pm, $_{63}$Eu and $_{65}$Tb~\cite{data2} are shown in 
Fig.~\ref{fig27}. 

\begin{figure}[h!]
\centering
\includegraphics[scale=0.42]{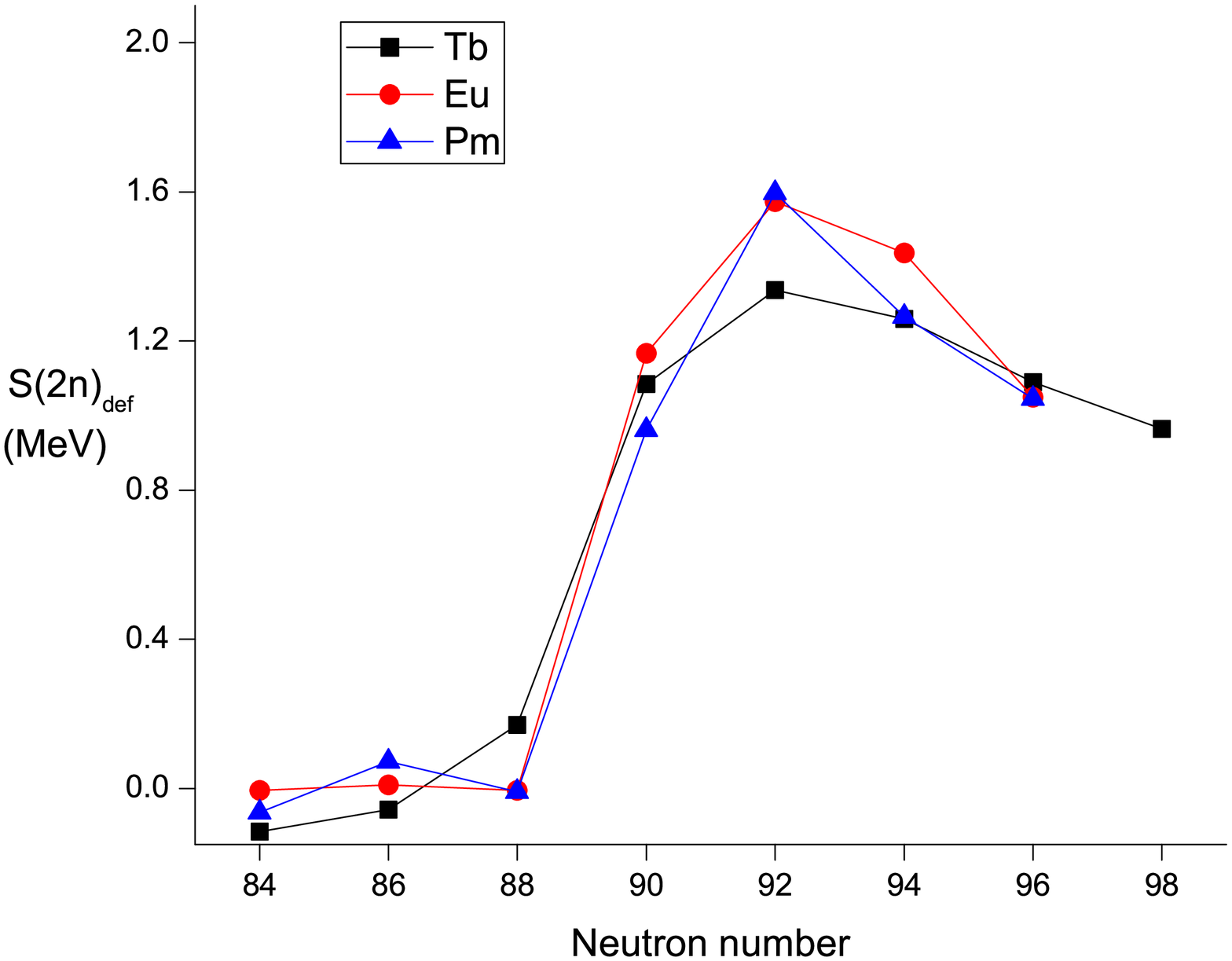}
\caption{The contribution of deformation, $S(2n)_{def}$ to the two-neutron
separation energies, Eq.~(\ref{Eq:S2n}), 
in $_{61}$Pm, $_{63}$Eu and $_{65}$Tb.
Experimental data taken from~\cite{data2}.}
\label{fig28}
\end{figure}

One can see clearly the occurrence of discontinuities in the behavior of 
the two-neutron separation energies, an indication of a QPT. In order to
emphasize these discontinuities, we note that the two neutron separation
energies are given by a smooth contribution linear in $N$ plus the
contribution of the deformation~\cite[p.74]{iac111}
\begin{equation}
S_{2n}=-A_{2n}-B_{2n}N+S_{2n}^{def} ~.
\label{Eq:S2n}
\end{equation}
In Fig.~\ref{fig28} we show the deformation contribution, 
extracted from the data by subtracting the
linear dependence, with $A_{2n}=-15.185,\, -16.37,\, -17.672$ MeV 
for Pm, Eu, Tb, respectively, and $B_{2n}=0.670$ MeV. 
\begin{figure}[h!]
\centering
\includegraphics[scale=0.42]{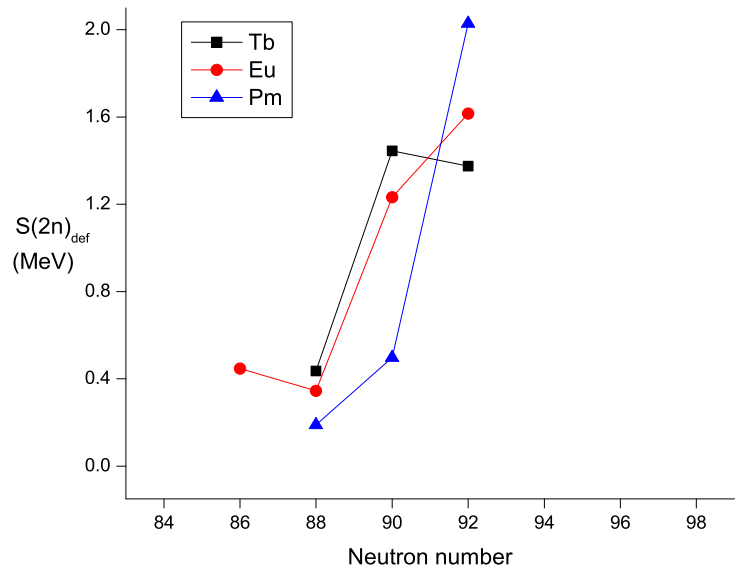}
\caption{The contribution of deformation, $S(2n)_{def}$ 
Eq.~(\ref{Eq:S2n}), calculated in IBFM with the parameters 
given in the text.}
\label{fig29}
\end{figure}
The deformation contribution can be easily calculated using IBFM. 
Fig.~\ref{fig29}
shows this contribution for the $h_{11/2}$ level, 
for $_{61}$Pm, $_{63}$Eu and $_{65}$Tb. 
Figs.~\ref{fig28} and~\ref{fig29} cannot be, in principle, 
directly compared since the
experimental data are the separation energies of the actual ground state,
while Fig.~\ref{fig29} shows the separation energies of the $h_{11/2}$ level.
However, in practice, differences between the two are of order $1/N$, and
thus, to a good approximation, Fig.~\ref{fig28} can be compared to 
Fig.~\ref{fig29}. In particular, the onset of deformation is clearly seen 
both in Fig.~\ref{fig28} and Fig.~\ref{fig29}.

\section{Conclusions}
\label{Sec6}

In this paper, we have analyzed QPTs in Bose-Fermi systems, specifically the
effect of one single-fermion with angular momentum~$j$ immersed in a bath of
bosons with angular momentum $L=0,2$ ($s,d$ bosons).

By doing a classical analysis (Section~\ref{Sec3}) we have studied the 
single particle motion in a field with $\beta ,\gamma $ deformation 
and determined the classical order parameters $\beta _{e,i},\gamma _{e,i}$. 
We have shown that 
while the presence of the odd-fermion does not influence much the motion in
the strongly deformed regions, the corrections being of order $1/N$, it does
influence greatly the location of the critical point and the entire nature
of the phase transition, washing out the transition for some states and
enhancing it for others. This is a novel result which may have applications
to other fields of physics outside of nuclear physics.

By doing a quantal analysis (Section~\ref{Sec4}) we have investigated 
the change in level structure induced by the phase transition 
(correlation diagrams). Here we have obtained the result that while the 
presence of the odd-nucleon does not affect much the 
ground-state energy ($1/N$ correction) it does affect
greatly the level structure. This level structure is rather complex,
especially in the neighborhood of the critical point. We have also used the
correlation diagrams to investigate the classical-quantal correspondence and
shown that when the bosons have SU(3) symmetry the correspondence is exact.

Finally, we have presented experimental evidence for the $U(5)-SU(3)$
transition (spherical to axially deformed) in odd-proton nuclei, $_{61}$Pm, $
_{63}$Eu and $_{65}$Tb, and performed a realistic calculation which accounts
well for the experimental level structure. This is in spite of the fact that
the structure of these nuclei is rather complex and cannot be simply
described either by the rotational or by the vibrational model.

Our investigation of the effects of a single fermion on the phase
transitions of a boson condensate has been done for the rather complex case
of bosons with angular momentum $L=0,2$ and fermion with angular momentum $
j=11/2$. It applies equally well to simpler cases, for example to the case
of a spin $j=1/2$ particle immersed in a bath of spinless bosons $L=0$. As
such, the method discussed here can be used in a variety of fields, ranging
from molecules to atomic condensates, from nuclei to mesoscopic systems.

\section*{Acknowledgements}

We thank C.E. Alonso, J.M. Arias, M. B\"oy\"ukata, M.A. Caprio, 
L. Fortunato and especially, A. Vitturi for many useful discussions on 
their work in QPT in Bose-Fermi systems. 
This work was performed in part under DOE\ Grant No. DE-FG-02-91ER40608 
and in part by a grant from the U.S.-Israel Binational Science Foundation.


\begin{thebibliography}{99}
\bibitem{gilmore1} 
R. Gilmore and D.H. Feng, 
Nucl. Phys. A 301 (1978) 189.

\bibitem{gilmore2} 
R. Gilmore, 
J. Math. Phys. 20 (1979) 891.

\bibitem{vojta} 
M. Vojta, 
Rep. Prog. Phys. 66 (2003) 2069.

\bibitem{diep} 
A.E.L. Dieperink, O. Scholten and F. Iachello, 
Phys. Rev. Lett. 44 (1980) 1747.

\bibitem{feng} 
D.H. Feng, R. Gilmore, and S.R. Deans, 
Phys. Rev. C 23 (1981) 1254.

\bibitem{iac-scholten} 
O. Scholten, F. Iachello and A. Arima, 
Ann. Phys. (N.Y.) 115 (1978) 325.

\bibitem{iac111} 
F. Iachello and A. Arima, 
\textit{The Interacting Boson Model} 
(Cambridge University Press, Cambridge, 1987).

\bibitem{cps1}
F. Iachello, 
Phys. Rev. Lett. 85 (2000) 3580.

\bibitem{cps2}
F. Iachello, 
Phys. Rev. Lett. 87 (2001) 052502. 

\bibitem{cps3}
F. Iachello, 
Phys. Rev. Lett. 91 (2003) 132502.

\bibitem{qds}
D.J. Rowe, 
Phys. Rev. Lett. 93 (2004) 47.

\bibitem{pds}
A. Leviatan, 
Phys. Rev. Lett. 98 (2007) 242502. 

\bibitem{exp1}
R.F. Casten, E.M. McCutchan, 
J. Phys. G: Nucl. Part. Phys. 34 (2007) R285. 

\bibitem{exp2}
R.F. Casten, 
Prog. Part. Nucl. Phys. 62 (2009) 183.

\bibitem{levgin03}
A. Leviatan and J.N. Ginocchio, 
Phys. Rev. Lett. 90 (2003) 212501.

\bibitem{zamfir} 
F. Iachello and N.V. Zamfir, 
Phys. Rev. Lett. 92 (2004) 212501.

\bibitem{lev05}
A. Leviatan, 
Phys. Rev. C 72 (2005) 031305(R).

\bibitem{lev06}
A. Leviatan, 
Phys. Rev. C 74 (2006) 051301(R).

\bibitem{rowe04}
D.J. Rowe, P.S. Turner and G. Rosensteel, 
Phys. Rev. Lett. 93 (2004) 232502. 

\bibitem{dusuel05} 
S. Dusuel, J. Vidal, J.M. Arias, J. Dukelsky and J.E. Garcia-Ramos, 
Phys. Rev. C 72 (2005) 011301(R).

\bibitem{williams}
E. Williams, 
{\it ``A study of transitional collective behaviour in heavy nuclei''}, 
Ph.D. Thesis, Yale University (2009).

\bibitem{werner} 
E. Williams, R.J. Casperson and V. Werner, 
Phys. Rev. C 81 (2010) 054306.

\bibitem{caprio} 
M.A. Caprio, P. Cejnar and F. Iachello, 
Ann. Phys. (N.Y.) 323 (2008) 1106.

\bibitem{cejnar09}
P. Cejnar and J. Jolie, 
Prog. Part. Nucl. Phys. 62 (2009) 210. 

\bibitem{cejnar10}
P. Cejnar, J. Jolie and R.F. Casten, 
Rev. Mod. Phys. 82 (2010) 2155.  

\bibitem{iaccap10}
F. Iachello and M.A. Caprio,
in {\it Understanding Quantum Phase Transitions}, L. Carr ed., Taylor and 
Francis (2010).

\bibitem{iac1} 
F. Iachello and P. Van Isacker, 
\textit{The Interacting Boson-Fermion Model} 
(Cambridge University Press, Cambridge, 1991).

\bibitem{lehur} 
K. Le Hur, 
in \textit{Understanding Quantum Phase 
Transitions}, L. Carr, ed., Taylor and Francis (2010).

\bibitem{scholten-blasi} 
O. Scholten and N. Blasi, 
Nucl. Phys. A 380 (1982) 509.

\bibitem{alonso1} 
C.E. Alonso, J.M. Arias, L. Fortunato and A. Vitturi,
Phys. Rev. C 72 (2005) 061302(R).

\bibitem{alonso2} 
C.E. Alonso, J.M. Arias, and A. Vitturi, 
Phys. Rev. C 75 (2007) 064316.

\bibitem{alonso3} 
C.E. Alonso, J.M. Arias, L. Fortunato and A. Vitturi,
Phys. Rev. C 79, 014306 (2009).

\bibitem{boy10}
M. B\"oy\"ukata, C.E. Alonso, J.M. Arias, L. Fortunato and A. Vitturi,
Phys. Rev. C 82, 014317 (2010).

\bibitem{reviews1} 
F. Iachello, 
Proc. Int. School "Enrico Fermi", Course
CLIII, A. Molinari, L. Riccati, W.M. Alberico, and M. Morando, eds. 
(IOS Press, Amsterdam, 2003).

\bibitem{landau} 
L. Landau and E.M. Lifshitz, 
{\it Statistical Physics}, (Pergamon, Oxford, 1980).

\bibitem{lev88} 
A. Leviatan, 
Phys. Lett. B 209 (1988) 415.

\bibitem{levshao89} 
Amiram Leviatan and Bin Shao, 
Phys. Rev. Lett. 63 (1989) 2204.

\bibitem{ginocchio} 
J.N. Ginocchio and M.W. Kirson, 
Phys. Rev. Lett. 44 (1980) 1744.

\bibitem{bohr} 
A. Bohr and B.R. Mottelson, 
Phys, Scr. 22 (1980) 468.

\bibitem{mayer} 
J. Meyer-ter-Vehn, 
Nucl. Phys. A 249 (1975) 111.

\bibitem{nilsson} 
S.G. Nilsson, 
K. Dan. Vidensk. Selsk. Mat.-Fys. Medd. 29, No 16 (1955).

\bibitem{scholtenodda} 
O. Scholten, 
computer program ODDA.

\bibitem{scholtenthesis} 
O. Scholten, 
\textit{The Interacting Boson
Approximation Model and Applications,} 
Ph.D. Thesis, University of Groningen, The Netherlands (1980).

\bibitem{scholten-ozzello} 
O. Scholten and T. Ozzello, 
Nucl. Phys. A 424 (1984) 221.

\bibitem{data1} 
Evaluated Nuclear Structure Data File (ENSDF), 
{\it http://www.nndc.bnl.gov/ensdf/browse\_top.jsp}.

\bibitem{data2} 
LBNL Isotopes Project Nuclear Data Dissemination Home Page, 
retrieved April 30, 2010,\\ 
{\it http://ie.lbl.gov/toi2003/MassSearch.asp}.

\end{thebibliography}
\end{document}